\documentclass[preprint,12pt]{elsarticle}
\usepackage{amssymb}
\usepackage{amsmath}

\journal{arXiv}

\usepackage{amsmath}
\usepackage{array}
\usepackage{url}
\usepackage{graphicx}
\usepackage{xcolor}
\usepackage{adjustbox,listings}
\usepackage{float}
\usepackage{orcidlink}

\begin{document}

\begin{frontmatter}

\title{Statistically Supported LLM Ingredient and Recipe Data Collection in Computational Nutrition}

\author[SU]{James Izzard\orcidlink{0009-0006-3916-4514}}
\ead{2377507@swansea.ac.uk}

\author[SU]{Hassan Eshkiki\orcidlink{0000-0001-7795-453X}}
\ead{h.g.eshkiki@swansea.ac.uk}

\author[SU]{Fabio Caraffini \corref{mycorrespondingauthor}\orcidlink{0000-0001-9199-7368}}
\cortext[mycorrespondingauthor]{Corresponding author}
\ead{fabio.caraffini@swansea.ac.uk}

\affiliation[SU]{organization={Swansea University},
            addressline={Department of Computer Science, Bay Campus},
            city={Swansea},
            postcode={SA1 8EN},
            state={Wales},
            country={UK}}

\begin{abstract}
Computational nutrition needs precise ingredient data, but current databases are incomplete, inconsistent, and built for human reference rather than automated reasoning. LLMs could help fill these gaps, but single-pass outputs are unreliable and can introduce silent errors into downstream computation. We present a quality-controlled LLM pipeline for ingredient data acquisition that combines robust statistical estimation, domain-specific invariant checks, and a web-fetch fallback. An illustrative Heap's Law fit to 233 recipes suggests that unique-ingredient growth is sub-linear and front-loaded: the projected ratio of unique ingredients to recipes falls from 1.74 at 100 recipes to 0.19 at 5,000. For each ingredient attribute, repeated LLM queries are treated as samples from a model-induced answer distribution, and we apply robust point estimators and normalised confidence scores across numerical, Boolean, multiple-choice, open categorical, and optional integer types. An invariant guard layer enforces nutritional and logical self-consistency within each ingredient record. Minor numeric inconsistencies are reconciled via a linear program that minimises worst-case percentage deviation while preserving semantic zeros, and major violations are escalated to web-evidence-grounded repair, then human review only if that fails. On a curated 30-ingredient reference set, the pipeline achieves 98.4\% exact match on nutrient flags and cuts median absolute percentage error on nutrient ratios from 31.9\% for the median-aggregated baseline to 10.1\%, a reduction of 21.8 percentage points, at an API cost of about \$1 per ingredient. This frames LLM-assisted database construction as a controlled data-engineering workflow that makes uncertainty operational rather than discarding it.

\end{abstract}


\end{frontmatter}

\section{Introduction}
\label{sec:introduction}

A computational nutrition system is only as useful as the data it reasons over. Nutritional planning and analysis require detailed, structured, and internally consistent ingredient records, yet existing databases are typically incomplete, inconsistent between sources, and structured for human browsing rather than computational use \cite{li2023,brinkley2025}. The renewed interest in applying large language models (LLMs) and AI to nutritional planning has only sharpened this bottleneck. Detailed data for thousands of ingredients are needed at a scale that manual curation cannot match, while cheap bulk methods reach that scale only by compromising quality, so neither delivers usable data.

LLMs offer a plausible route to build such a source. They carry extensive implicit knowledge of ingredients and their compositions, and recent work has shown that they can populate structured scientific databases at scale with accuracies comparable to human annotators, including in the nutrition domain specifically \cite{dagdelen2024,zheng2023,hua2025}. The immediate difficulty is that individual LLM outputs are unreliable in a way that is especially awkward for this application: a single confidently asserted, but wrong, nutrient value silently corrupts every downstream computation.

Simply asking the same question many times and averaging does not resolve this. The classical wisdom-of-crowds argument depends on errors being independent across estimators, but repeated samples from the same model share its parameters, training distribution, and decoding process, so errors are correlated \cite{denisovblanch2026}. Nevertheless, previous work on repeated sampling and consistency-based uncertainty estimation \cite{wang2023,farquhar2024} suggests that disagreement among generated responses carries useful information. We use this observation, but treat dispersion as an indicator of model confidence rather than as proof of correctness: tight agreement reflects stable model belief but does not guarantee accuracy, so external checks remain necessary.

A second tempting strategy is to ask the model to repair its own contradictions once detected. This produces numerically consistent records without grounding them. The constraint is underdetermined, and in our experience models tend to satisfy it the cheap way, reaching for a numerically consistent arrangement over an accurate one even when asked to ground their values in evidence. The method proposed here, therefore, separates elicitation from validation. Each field is sampled independently to estimate the model-induced answer distribution, and external checks for numerical and semantic consistency are applied only afterwards. Where these checks fail, repair is grounded in fresh evidence retrieved by automated web search rather than only self-correction prompts.

In this study, we develop the methodology around that separation. The statistical layer estimates each field by repeated sampling: numerical values use the median and median absolute deviation, and Boolean, multiple-choice, open categorical and optional integer fields use analogous robust estimators with normalised agreement measures and stopping criteria. An invariant-checking layer enforces domain-specific constraints, for example by ensuring that the quantitative values associated with child nutrients aggregate to the value specified for their parent nutrient. The dispersion-based filtering catches noisy errors, while invariant checking catches systematic errors that survive aggregation.

Viewed more broadly, these layers connect the present work to information fusion under imperfect evidence. Information-fusion systems must combine inputs affected by uncertainty, incompleteness, outliers, conflict and source dependence \cite{hall1997,khaleghi2013}. The same problems arise here, although the inputs are not conventional sensor measurements. Instead, repeated LLM responses provide correlated observations from one generative source, web retrieval supplies external factual evidence, and nutritional invariants contribute structural domain knowledge. The pipeline preserves these distinct evidential roles rather than treating every input as an interchangeable vote, using each to estimate, ground or constrain the resulting ingredient record.

While the invariant layer depends on nutrition-specific structure, the statistical layer is source-agnostic and may transfer to any domain where structured records must be assembled from noisy generative estimates.

The article is organised as follows. Section~\ref{sec:related} reviews related work, Section~\ref{sec:heaps} projects the scale of the collection task using a Heap's Law model, and Section~\ref{sec:model-pipeline} defines the ingredient record and full collection pipeline. Section~\ref{sec:repeat-queries} presents the per-data-type estimation framework, Section~\ref{sec:experimental-design} defines the experimental design, and Section~\ref{sec:saturation} reports the saturation-sampling study. Section~\ref{sec:invariants} develops the invariant-checking layer with its linear-programming formulation for nutrient-tree balancing. Section~\ref{sec:results} presents the head-to-head comparison. Sections~\ref{sec:discussion} and~\ref{sec:conclusion} discuss limitations and conclude.

\section{Literature Review}
\label{sec:related}

Our contribution sits at the intersection of four threads: the inadequacy of existing nutritional databases, the emerging capability of LLMs to populate structured scientific databases, the unreliability of individual LLM outputs, and the statistical machinery needed to convert noisy samples into trustworthy data.

\subsection{Existing Nutritional Databases Are Insufficient for Computational Use}
\label{sec:related-databases}

Coverage of existing nutritional databases is narrow and uneven. Barabasi et al. \cite{barabasi2020} estimate that standard food composition databases track roughly 150 of the more than 26,000 biochemicals present in food, terming the remainder ``nutritional dark matter''. Even within currently recognised nutrients, coverage remains incomplete. Li et al. \cite{li2023}, evaluating 175 food and nutrient sources, found that even USDA SR Legacy is incomplete for all 40 essential nutrients identified by the US National Academies of Sciences, Engineering, and Medicine, with per-food-group completeness averaging around 70\%.

Brinkley et al. \cite{brinkley2025} found that database quality is stratified by national income, where high-income countries dominate the supply of primary analytical data, regular updates, and FAIR-compliant databases, leaving lower-income countries reliant on secondary data that can misrepresent their local food supply.

Internal inconsistency compounds the problem: Ferraz de Arruda et al. \cite{ferrazdearruda2023} identified entries in the Spanish BEDCA database whose fatty-acid values summed to over 100 g per 100 g, a physical impossibility. Van Erp et al. \cite{vanerp2021} observe that nutrition and recipe databases are typically not cross-referenced, with ingredient quantities and preparation states expressed inconsistently across sources. Any practical collection process must therefore be extensible as dietary science advances.

This database bottleneck also appears in general meal-optimisation software. Izzard et al. \cite{bib:izzard2023} showed that a flexible ingredient-recipe model can support diverse nutritional, medical, cultural, and economic constraints, but also identified reusable, validated ingredient and recipe data as a long-term requirement for such systems. The present work addresses that unresolved data-population problem by treating ingredient records as objects to be estimated, checked, and repaired before use in downstream optimisation.

\subsection{LLMs Can Support Structured Scientific and Nutrition Data Tasks}
\label{sec:related-llm-data-tasks}

Recent work suggests that LLMs can extract structured records from unstructured scientific text with usable accuracy. Dagdelen et al. \cite{dagdelen2024} fine-tuned LLMs on around 500 annotated examples to perform joint named entity recognition and relation extraction across three materials chemistry tasks with high precision. Closer to our prompting-only setting, Zheng et al. \cite{zheng2023} used ChatGPT to extract 26,257 synthesis parameters from roughly 800 metal-organic framework papers with F1 between 90\% and 99\%, and the resulting database was sufficient to train downstream models predicting crystallisation outcomes with over 87\% accuracy.

Nutrition-specific work shows a related pattern. Bedrac et al. \cite{bedrac2025} show Llama-3 70B achieving F1 of 0.894 on decomposing recipes into ingredients mapped to USDA FoodData Central, a task close in spirit to our own. On the harder task of estimating nutrition directly from meal descriptions, Hua et al. \cite{hua2025} report GPT-4o reaching 66.8\% accuracy within a 7.5 g carbohydrate tolerance on their NutriBench benchmark, outperforming human nutritionists on both speed and accuracy but mapping each description straight to a macronutrient total, with no intermediate ingredient breakdown against which a result can be checked.

Gjorgjevikj et al. \cite{GJORGJEVIKJ2026101351} introduce FoodyLLM, a domain-specialised model fine-tuned for food and nutrition tasks such as nutrient estimation, traffic-light classification, and food-entity linking. This demonstrates the value of adapting LLMs to food-specific representations, but it also illustrates a different architectural choice: the LLM remains the component performing the user-facing task. The present work instead uses the LLM only to construct ingredient records, through confidence estimation, invariant checking, and fallback retrieval, which are then consumed by deterministic computational nutrition models. The generative cost is therefore incurred once, at collection time, and amortised across every subsequent query, whereas an LLM in the user-facing layer pays a token cost on every question asked.

Vavken et al. \cite{Vavken_2025} evaluate LLMs as query interpreters in a food-composition RAG system, finding that metadata filtering performs well when user constraints can be explicitly represented, but becomes less reliable when queries exceed the representational scope of the metadata.

Together, these studies indicate that LLMs can make food and nutrition data more accessible and computationally usable, yet the approaches explored to date lack the statistical rigour and internal consistency checks that the present work introduces.

\subsection{Individual LLM Outputs Are Unreliable, But Their Dispersion Is Informative}
\label{sec:related-dispersion}

The counterweight is well-documented hallucination \cite{huang2025}. Usefully, this unreliability is often detectable. When the same question is asked repeatedly at non-zero temperature, responses well-supported by training data cluster tightly, while guessed or unsupported responses scatter. This signal admits two uses. It can sharpen answers, as when Wang et al. \cite{wang2023} take the majority over sampled chain-of-thought paths, improving accuracy by up to 18 points on reasoning benchmarks. It can also flag unreliable ones, as when Farquhar et al. \cite{farquhar2024} formalise semantic entropy, where hallucination is signalled by generations falling into many incompatible meaning clusters rather than one stable answer. Geng et al.'s survey \cite{geng2024} identifies repeated sampling with consistency measurement as an established black-box method.

Repeated sampling does not yield statistically independent error terms, however. Every sample comes from the same model weights and training distribution, so any systematic mistake is shared between all draws rather than averaged away. Denisov-Blanch et al. \cite{denisovblanch2026} show that polling-style aggregation does not provide a reliable signal of truthfulness, because errors remain correlated between samples. The consensus reflects the belief of the shared model, not the shared evidence. Distinguishing the two requires a grounding external to the model. We therefore treat dispersion not as an estimator but as an initial confidence gate; low-dispersion items pass forward into domain validation, while high-dispersion items are flagged rather than propagated as known.

\subsection{The Median Absolute Deviation Provides A Robust Confidence Metric}
\label{sec:related-mad}

Standard deviation is a poor dispersion measure for repeated LLM estimates because a single extreme response can dominate it; its breakdown point is zero. The median absolute deviation instead tolerates up to half the sample being corrupted before it loses meaning, a 50\% breakdown point \cite{rousseeuw1993}. Rousseeuw and Croux \cite{rousseeuw1993} proposed alternatives with higher Gaussian efficiency, but none improves on this breakdown point, and MAD's simplicity and interpretability have kept it the default in applied robust statistics. The structure of repeated LLM sampling, a central cluster of well-supported responses surrounded by sparser unsupported outliers, is exactly the contamination regime MAD was designed for.

\subsection{Synthesis}
\label{sec:related-synthesis}

The distinctive move of this work is to treat ingredient characterisation as a data-engineering task separated from nutritional reasoning: the expensive generative step is spent once, on building grounded, internally consistent records, rather than repeated at every downstream query. No prior work assembles repeated-sampling estimation, invariant checking, and external grounding into a single ingredient-collection pipeline of this kind. The closest methodological neighbour for the uncertainty component alone is the semantic entropy of Farquhar et al. \cite{farquhar2024}, but that was developed for free-form question answering rather than the structured numerical fields treated here. Our contribution is to demonstrate empirically that the combined approach yields ingredient records of materially higher quality than naive one-shot collection, at a cost suited to a one-time corpus build.

\section{Modelling Number of Ingredients vs Number of Recipes}
\label{sec:heaps}

Before embarking on large-scale data collection, it is useful to understand the scope and characteristics of the task. Specifically, we can estimate how many unique ingredients must be characterised to support a database of a given number of recipes. This estimate informs expectations for the data collection effort.

The naive assumption might be that the number of unique ingredients grows linearly with the number of recipes. If each recipe introduces entirely new ingredients, a database of 1,000 recipes with an average of 10 ingredients per recipe would require 10,000 unique ingredient entries. This assumption is clearly incorrect, since recipes extensively share ingredients; onions, garlic, olive oil, and salt appear across cuisines and recipe categories, meaning that each additional recipe contributes progressively fewer new ingredients to the database.

This property, where the number of unique elements grows sub-linearly with the total number of element occurrences, is well characterised by Heap's Law \cite{heaps1978}.

\subsection{Heap's Law}
\label{sec:heaps-law}

Heap's Law is an empirical relationship originally observed in computational linguistics that describes how the number of unique words (vocabulary size) grows as a function of the total number of words in a corpus. The law states that vocabulary grows according to a power law with an exponent less than one, meaning that larger corpora exhibit diminishing returns in terms of new vocabulary.

The standard formulation is:

\begin{equation}
V = K \cdot N^\beta
\end{equation}

$V$ is the vocabulary size (number of unique words) $N$ is the corpus size, measured as word tokens: total word occurrences rather than unique words $K$ is a scaling constant $\beta$ is the growth exponent, typically $0.4 \leq \beta \leq 0.6$.

Heap's Law has since been observed beyond linguistics, wherever elements are drawn from a long-tailed distribution. Benz et al. \cite{benz2008} established Heap-like growth for unique chemical substructures in molecular databases, a close conceptual analogue to our own setting; chemical features and food ingredients are drawn from a large but not unbounded pool, with a few common elements appearing frequently and many rare ones appearing rarely. Tria et al. \cite{tria2018} provide a unified ``adjacent possible'' framework showing that Heaps-like growth emerges in systems where each observed element expands the set of plausible future elements, across domains as diverse as ecological systems, social tagging, and technological innovation.

The sub-linear growth ($\beta < 1$) reflects the fact that common words are reused frequently, whilst rare words appear infrequently.

\subsection{Adaptation To Recipe-Ingredient Modelling}
\label{sec:heaps-recipe-ingredient}

We adapt Heap's Law to model the relationship between recipes and unique ingredients. The corpus size $N$ becomes the total number of ingredient occurrences across all recipes, which equals the number of recipes multiplied by the average number of ingredients per recipe. This gives:

\begin{equation}
I = c \cdot (k \cdot N_r)^\beta,
\end{equation}

where $I$ is the number of unique ingredients, $N_r$ is the number of recipes, $k$ is the average number of ingredients per recipe, $c$ is a scaling constant, $\beta$ is the growth exponent.

\subsection{Parameter Estimation}
\label{sec:heaps-parameter-estimation}

Our aim is to obtain a rough sense of how the unique-ingredient count grows with the number of recipes, so we can anticipate the order of magnitude of the data collection effort.

We fitted the model to an existing dataset of 233 recipes. The recipes were placed in a single random order, and the cumulative count of unique ingredients was recorded at 17 nested sample sizes spanning from 5 to 233 recipes. The average ingredients per recipe $k$ was measured directly over the full set and held fixed, so that the corpus size $k \cdot N_r$ is formed with the same $k$ at every sample point. The growth exponent $\beta$ and scaling constant $c$ were then obtained by ordinary least squares on the log-transformed relation $\ln I = \beta \ln(k \cdot N_r) + \ln c$.

The fitted parameters and their values are reported in Table \ref{tab:heaps-params}.

\begin{table}[ht!]
\caption{Fitted Heap's Law Parameters}
\label{tab:heaps-params}
\centering
\begin{tabular}{lll}
\hline
\textbf{Parameter} & \textbf{Value} & \textbf{Interpretation}\\
\hline
$k$ & 10.08 & Average ingredients per recipe (measured directly)\\
$\beta$ & 0.421 & Growth exponent\\
$c$ & 9.99 & Scaling constant\\
$R^2$ & 0.974 & Coefficient of determination\\
\hline
\end{tabular}
\end{table}

The fitted exponent is $\beta = 0.421$, indicating growth is sub-linear; each additional recipe contributes progressively fewer new ingredients. The value falls within the typical range observed in linguistic applications, suggesting ingredient reuse follows broadly similar statistical regularities to word reuse in natural language.

The power-law form fits the sampled accumulation closely, with $R^2 = 0.974$ on the log-log regression. This indicates that the chosen shape describes the observed curve well. It is not a measure of predictive accuracy on unseen data, as the fit is in-sample and the sampled points form a nested cumulative sequence rather than independent observations, but for anticipating collection cost a good description of the shape is sufficient.

\subsection{Projections}
\label{sec:heaps-projections}

Using the fitted model, we can project the number of unique ingredients required for databases of various sizes (Table \ref{tab:heaps-projections}). The diminishing Ingredients Per Recipe Ratio (IPRR) illustrates the practical implication of sub-linear growth: whilst a 500-recipe database averages approximately 0.72 unique ingredients per recipe, a 5,000-recipe database averages only 0.19. This has significant implications for data collection effort.

\begin{table}[ht!]
\renewcommand{\arraystretch}{1.2}
\caption{Projected ingredient counts by recipe-database size.}
\label{tab:heaps-projections}
\centering
\begin{tabular}{lll}
\hline
\textbf{Recipes} & \textbf{Predicted Unique Ingredients} & \textbf{IPRR}\\
\hline
100 & 174 & 1.74\\
250 & 267 & 1.07\\
500 & 361 & 0.72\\
1,000 & 483 & 0.48\\
2,000 & 647 & 0.32\\
5,000 & 951 & 0.19\\
\hline
\end{tabular}
\end{table}

Characterising the ingredient database is a front-loaded task. The cost of each additional recipe decreases substantially as the database grows. These projections extrapolate well beyond the fitted range of 233 recipes. However, the qualitative conclusion of sub-linear, front-loaded growth is robust to plausible variation in $\beta$.

\begin{figure}[ht!]
\centering
\includegraphics[width=\linewidth]{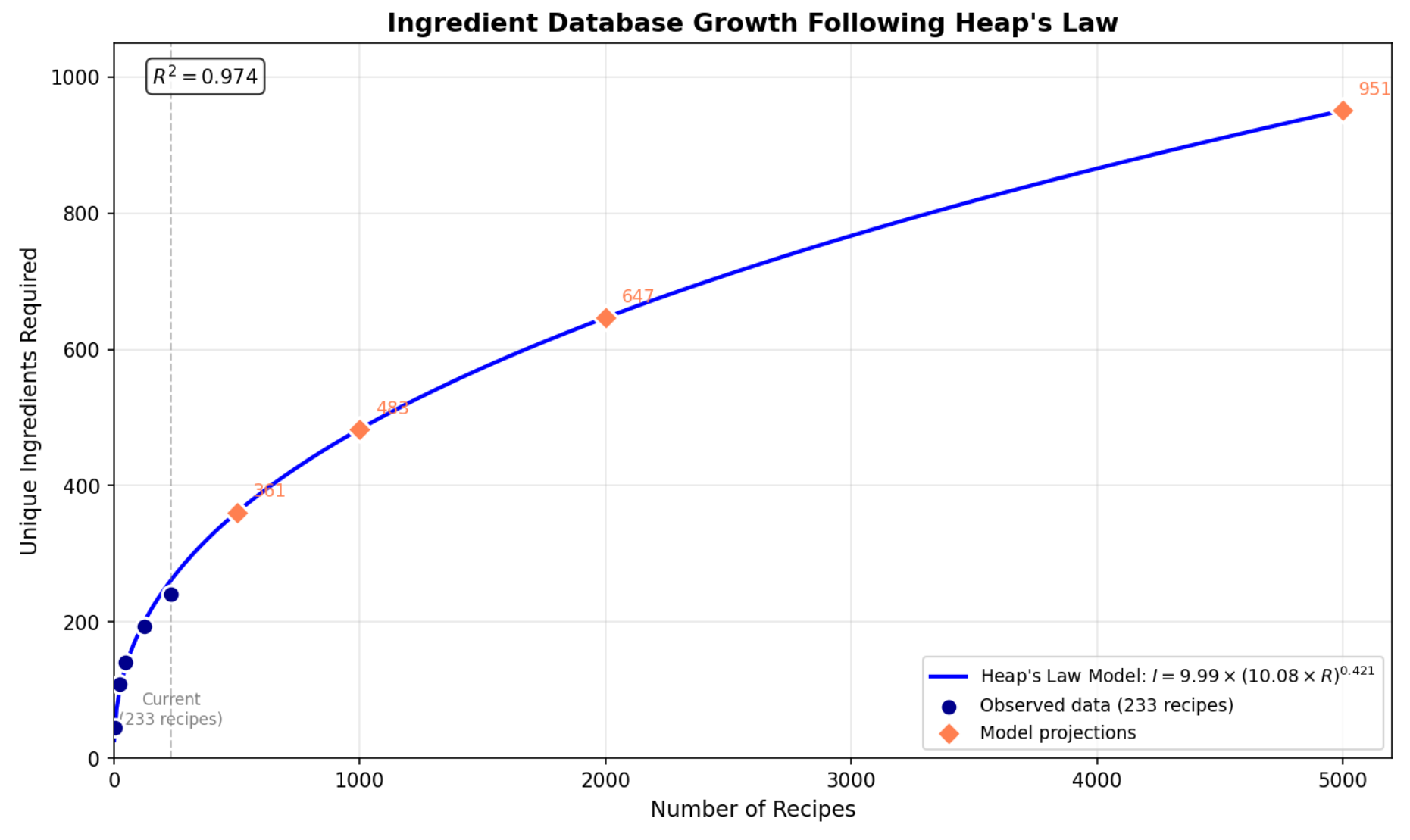}
\caption{Heap's Law fit to observed ingredient counts, projected to 5,000 recipes.}
\label{fig:heaps-fit}
\end{figure}

Figure \ref{fig:heaps-fit} illustrates the model fit to the observed data and the projection to 5,000 recipes.

\section{Data Model and Collection Pipeline}
\label{sec:model-pipeline}

We treat ingredient characterisation as an information extraction task that takes an ingredient name as input and produces a structured ingredient record. We refer to such an input as a candidate ingredient: a raw ingredient name, obtained either via automatic recipe decomposition or direct user entry, that has not yet been resolved to a record. The processes by which these candidates are obtained are outside the scope of the present evaluation, since recipes admit multiple valid decomposition schemes with no single, universally accepted canonical representation.

\subsection{Collection Pipeline}
\label{sec:collection-pipeline}

A candidate ingredient is first checked against the existing database, to prevent duplicate collections. New candidates have their reference quantity fetched first, because the reference quantity is used in subsequent prompts. The remaining fields (cost, glycaemic index, nutrient flags and ratios) are then obtained by the repeated-sampling estimators of Section~\ref{sec:repeat-queries}, with a web-fetch fallback for fields the sampler cannot resolve. 

Mandatory fields that web-fetch also cannot resolve are flagged for manual review. The assembled record is then checked against the invariants of Section~\ref{sec:invariants}. Where a check fails and a deterministic repair exists, it is applied; otherwise an LLM repair proposal is requested with the specific failure context, and the record is rechecked. If repair cannot resolve the failure within the retry budget, the record is escalated to manual review.

\begin{figure}[ht!]
\centering
\includegraphics[width=0.6\linewidth]{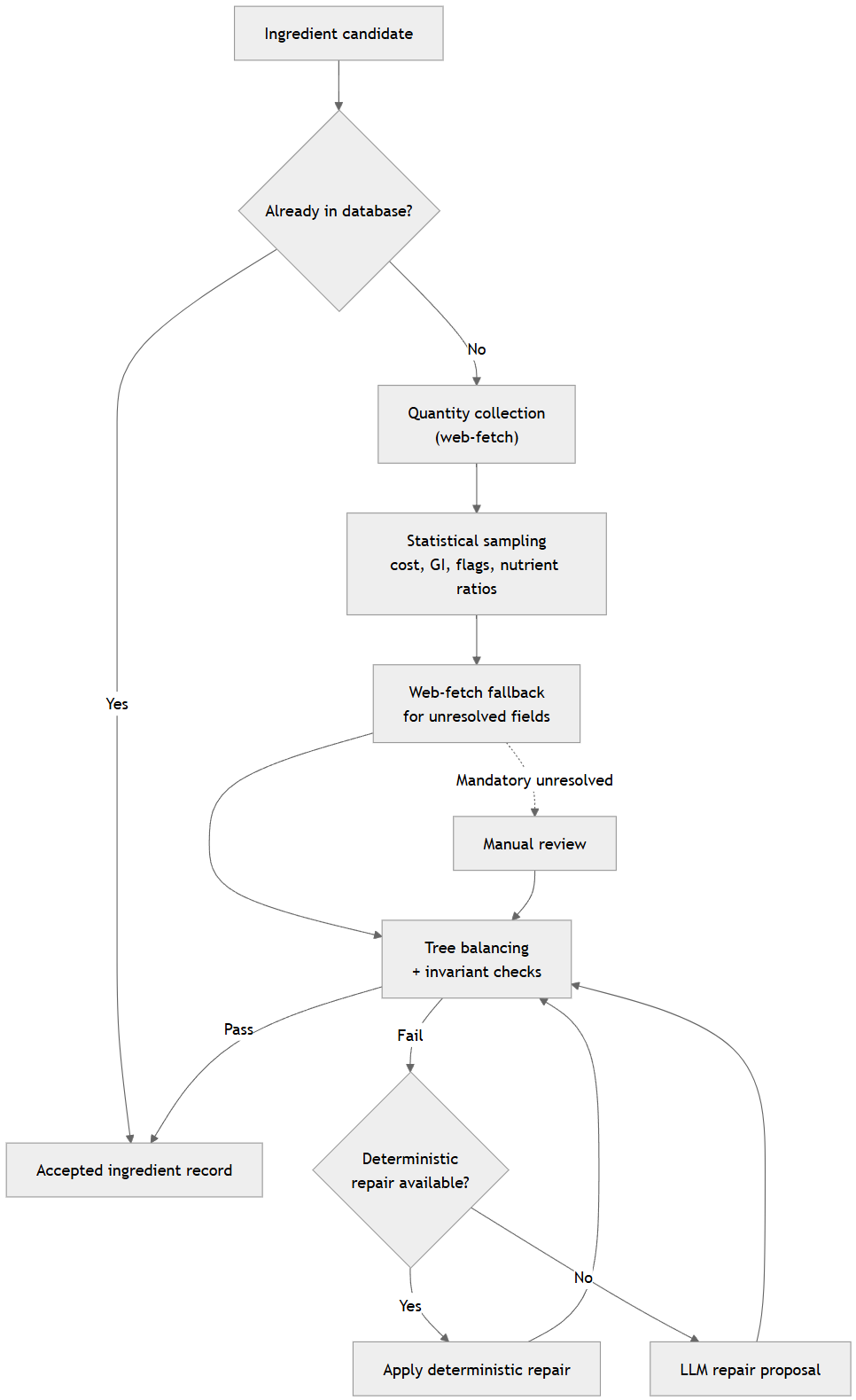}
\caption{Ingredient-characterisation pipeline.}
\label{fig:pipeline}
\end{figure}

Figure~\ref{fig:pipeline} summarises how a candidate ingredient moves through the collection pipeline.

Each pipeline stage contributes a different kind of information. Statistical sampling supplies the well-known fields cheaply, web-fetch handles less certain fields at higher per-call cost, and the invariant layer brings no new data but reconciles whatever the first two produce. None is sufficient alone; together they yield records that are relatively inexpensive at scale and self-consistent.

\subsection{Ingredient Data Structure}
\label{sec:ingredient-data-structure}

Each ingredient record contains a fixed set of attributes (Table \ref{tab:ingredient-attributes}).

\begin{table}[ht!]
\caption{Ingredient-record attributes collected by the pipeline.}
\label{tab:ingredient-attributes}
\centering
\begin{adjustbox}{max width = \textwidth}
\begin{tabular}{lp{5.5cm}p{4cm}}
\hline
\textbf{Field} & \textbf{Role} & \textbf{Example}\\
\hline
\texttt{reference\_quantity} & Natural unit for user-facing quantities; magnitude is implicit & whole (apple), litre (milk), slice (bread)\\
\texttt{grams\_per\_ref\_quantity} & Mass in grams of one reference quantity, anchoring all ratios & 100 ml milk is approximately 100 g\\
\texttt{cost\_per\_ref\_quantity} & Price per reference quantity & 2.50 GBP per kg rice\\
\texttt{glycaemic\_index} & Glycaemic response; partly constrained by nutrient ratios & GI 68\\
\texttt{nutrient\_ratios} & Nutrient mass per reference quantity & 75 g carbohydrate per 100 g flour\\
\texttt{nutrient\_flags} & Binary dietary or compositional properties & alcohol-free, halal\\
\hline
\end{tabular}
\end{adjustbox}
\end{table}

\texttt{nutrient\_flags} are collected even when they are partially constrained by nutrient ratios. This deliberate redundancy allows for a domain consistency check. For example, an ingredient marked alcohol-free should agree with a zero alcohol entry in \texttt{nutrient\_ratios}, while a non-zero alcohol ratio should imply a false alcohol-free flag. Other flags, such as \emph{halal}, may encode rules that are not reducible to composition alone.

\section{Estimating Values From Repeated Queries}
\label{sec:repeat-queries}

We estimate each data-point by repeatedly querying the LLM and deriving both a point estimate and a confidence score from the spread of responses. Consistent responses suggest well-supported knowledge. Scattered responses indicate an absent or contradictory training signal.

Five distinct data types arise in ingredient characterisation: \emph{numerical} values such as normalised \texttt{NutrientRatio} values and costs; \emph{Boolean} values like \texttt{NutrientFlag} attributes, such as alcohol-free; \emph{multiple-choice} values like the unit of a reference quantity, chosen from a fixed set of allowed units; \emph{open categorical} values such as a reference quantity as a whole, drawn from an unbounded space; and \emph{optional integer} values such as the glycaemic index. The multiple-choice and open categorical types meet in the reference quantity, which is a compound object rather than a simple number. The value is something like \texttt{(100, grams)} or \texttt{(1, whole)}, pairing an unbounded numeric magnitude with a unit drawn from a fixed allowed set. The fixed-set unit is the multiple-choice part; the pair as a whole, with its unbounded magnitude, is what makes the reference quantity open categorical.

We treat all five within a single framework. For each data type, we define a robust point estimator, a dimensionless confidence score, and a threshold-based stopping criterion. The estimators differ in form but share the same robustness intuition: a central cluster of consistent responses should not be displaced by scattered bad responses unless the bad responses become numerous enough to dominate the sample. This uniformity allows the collection pipeline to apply the same query-measure-stop logic to heterogeneous fields.

The relationship with aggregation of the wisdom of crowds and the reasons we use dispersion as a confidence gate rather than as an aggregation strategy are discussed in Section \ref{sec:model-pipeline}. Empirical evidence for the assumptions of the framework appears in Section \ref{sec:results}.

\subsection{The Role of Temperature and Dispersion}
\label{sec:temperature-dispersion}

The approach outlined above yields information only if repeated queries can genuinely disagree, and whether they can depends on how the LLM is sampled. Deterministic decoding collapses repeated calls toward the same answer, so the dispersion signal disappears. Excessive sampling randomness can drown the model's useful signal in artificial noise. The useful regime is between these extremes where responses can vary, but the variation still reflects alternatives to which the model assigns non-negligible probability.

Non-zero sampling does not create knowledge from randomness; it exposes the model's uncertainty structure. Tightly clustered samples indicate a stable answer distribution, while multiple clusters or wide scatter signal uncertainty or contradictory support. If the model judges the question unanswerable, it can return the \texttt{DECLINE} sentinel defined below.

This gives the framework a concrete probabilistic reading. Repeated calls are samples of the response distribution induced by the prompt, model, and sampling settings. The point estimator describes the distribution's empirical location or dominant response, while the confidence score describes how tightly responses cluster around that estimate. A stopping criterion such as $C \geq \tau$ says that observed responses are stable enough to commit to a value; items that never satisfy it are routed onwards rather than treated as known.

The requirement is therefore not a particular temperature value, but a sampling regime that is neither deterministic nor excessively noisy. On gpt-5-mini this is the provider default: temperature is fixed at 1 and cannot be set through the API (Section~\ref{sec:model-prompting}). The consequences of that constraint are taken up in Section~\ref{sec:limitations}.

\subsection{Non-substantive Responses}
\label{sec:non-substantive-responses}

The LLM may return two kinds of non-substantive response, each handled differently from an ordinary erroneous one.

The first is a \texttt{DECLINE} keyword, returned when the model judges itself unable to answer. This is structurally different from an outlier. An outlier participates in estimation and is handled by estimator robustness, whereas a \texttt{DECLINE} carries no value and is set aside.

The second is a \emph{parse failure}, where the model attempts to answer but produces output that violates the expected schema or falls outside domain-enforced bounds, for example a glycaemic index of 250. Parse failures are also set aside, but counted separately from declines because their diagnostic content differs. Repeated declines indicate the model judges its own knowledge insufficient, whereas repeated parse failures indicate a systematic prompting or schema problem that additional querying is unlikely to resolve.

Both are removed before estimation. If $n$ queries yield $d$ declines and $f$ parse failures, the number of substantive parsed responses is $n_{s} = n - d - f$. The minimum-sample requirement applies only to $n_{s}$.

\begin{figure}[ht!]
\centering
\includegraphics[width=0.6\linewidth]{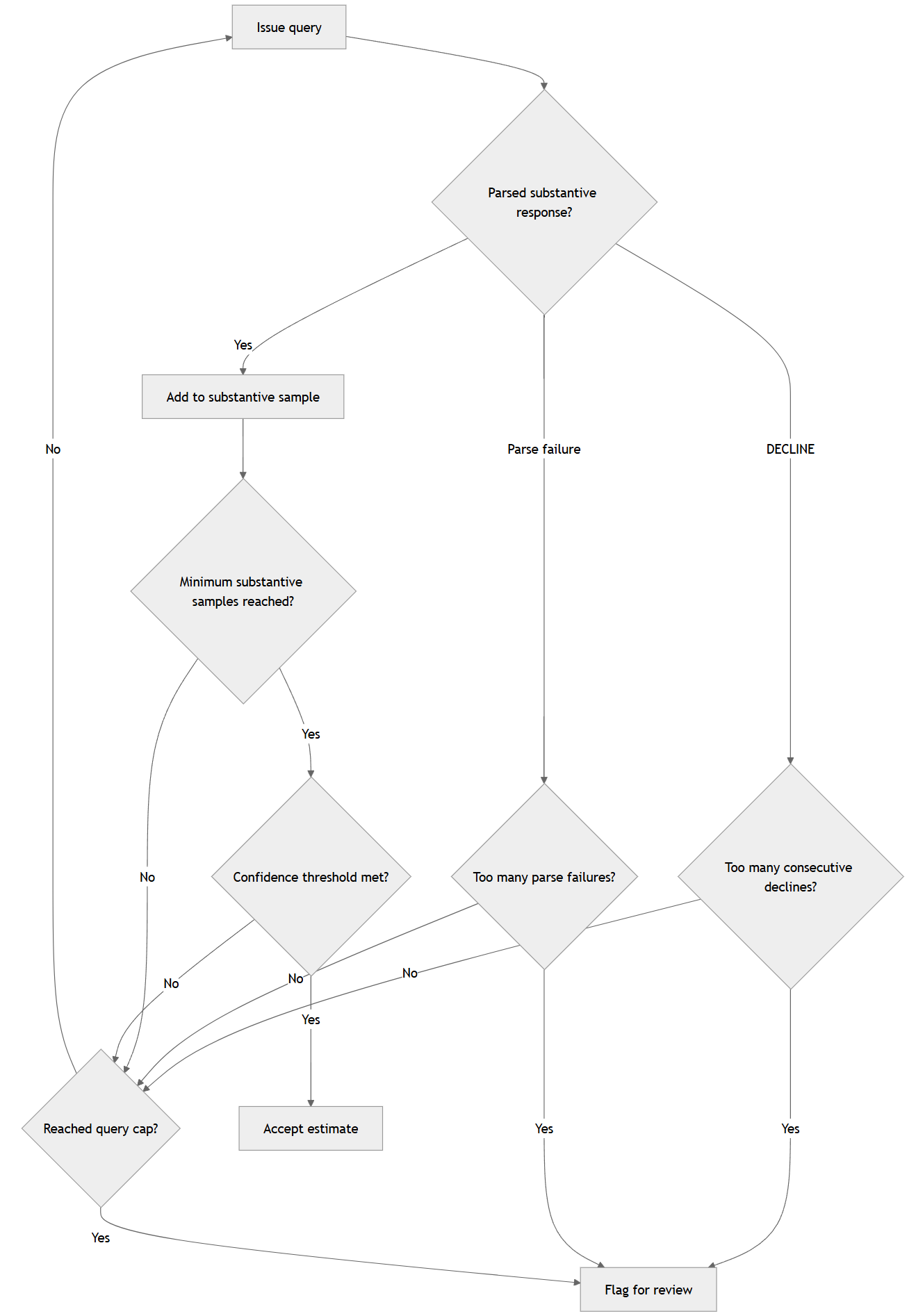}
\caption{Stopping logic for repeated sampling.}
\label{fig:stopping-logic}
\end{figure}

The resulting stopping logic is shown in Figure~\ref{fig:stopping-logic}.

In our experiments, we use $n_{\min} = 5$, $n_{\max} = 20$, a maximum of five consecutive declines, and a maximum of three parse failures. The value $n_{\min}=5$ is a pragmatic floor before confidence scoring is allowed, later supported by the saturation study's finding that rows that converge often do so as soon as this floor is reached. It is not a claim that all data types require identical evidence. The data-type-specific confidence score and the maximum-query cap determine how much additional sampling occurs.

\subsection{Numerical Values}
\label{sec:numerical-values}

Numerical values exhibit the clearest requirement for a robust estimator. A query for the protein content of chicken breast might return $31g$, $29g$, $32g$, $30g$, and $280g$ per $100g$ in five attempts. The arithmetic mean of $80.4g$ is more wrong than any individual response except the outlier itself, because it weights each observation by magnitude and so incorporates the outlier at face value. Therefore, an estimator that tolerates a proportion of arbitrarily corrupted responses is required.

The \emph{median} has this property. It depends only on the rank of observations, not their magnitude. Replacing the $280g$ outlier with $2800g$ or $28000 g$ leaves the median unchanged. Its breakdown point is 50\%, meaning the estimate remains anchored to the central cluster unless half or more of the responses are corrupted. We adopt the median $\tilde{x}$ as the point estimate.

To quantify confidence, we need a measure of spread with the same robustness. The standard deviation is inappropriate because it inherits the mean's sensitivity to outliers. We use the \emph{median absolute deviation} (Eq. \ref{eq:MAD}),

\begin{equation}\label{eq:MAD}
\text{MAD} = \text{median}(|x_i - \tilde{x}|)
\end{equation}

where double application of the median ensures robustness at both stages. The chosen reference point exhibits robustness to the influence of outliers, and the subsequent aggregation of deviations is similarly resistant to their effect. For the protein example, the absolute deviations from the median of $31 g$ are 2, 1, 0, 1, 249, giving a MAD of $1 g$. The outlier has negligible influence.

The MAD carries the units of the original data, so cross-nutrient comparison requires a dimensionless quantity. We use the \emph{robust coefficient of variation} (Eq. \ref{eq:CV}),

\begin{equation}\label{eq:CV}
\text{CV}_{\text{robust}} = \frac{1.4826 \cdot \text{MAD}}{|\tilde{x}|},
\end{equation}

where the constant $1.4826$ rescales the MAD such that, under the assumption of Gaussian-distributed errors, it becomes directly comparable to the standard deviation. Importantly, this constant does not impose a requirement of Gaussianity; it simply expresses the MAD-based measure on a conventional standard-deviation scale. Appendix \ref{app:mad} provides a concise explanation of this scaling. For the stopping criterion, we transform the robust coefficient of variation (CV) statistic into a bounded numerical confidence score (Eq. \ref{eq:Cn}),

\begin{equation}\label{eq:Cn}
C_n = \frac{1}{1 + \text{CV}_{\text{robust}}},
\end{equation}

with two special cases: unanimous zero responses receive a confidence of 1, while a zero median alongside non-zero responses receives a confidence of 0, since the spread cannot be expressed relative to a zero centre and the disagreement itself signals low confidence. The collection stops when $C_n \geq \tau$. In our experiments, we use a common acceptance threshold of $\tau = 0.90$, corresponding to $\text{CV}_{\text{robust}} \leq 0.111\ldots$ for numerical fields.

The framework is based on the assumption that consistency is a proxy for correctness. This assumption is not bulletproof. The LLM could confidently hallucinate the same wrong value every time, in which case any amount of dispersion-based filtering would produce false confidence. This consideration underscores the importance of the invariant-checking layer introduced in Section~\ref{sec:invariants}. This layer furnishes a complementary line of defence by validating responses against domain-specific logical constraints and thereby identifying systematic errors that may persist even after aggregation.

\subsection{Boolean Values}
\label{sec:boolean-values}

Boolean values are the special case $k = 2$ of the categorical family. The LLM returns \texttt{True} or \texttt{False} (or \texttt{DECLINE}), and the point estimate is the majority vote. To measure confidence, we begin with the raw agreement ratio $p_{\text{maj}} = n_{\text{maj}}/n_s$, where $n_{\text{maj}}$ is the number of responses matching the majority value and $n_s$ is the number of substantive parsed responses, and normalise against a chance baseline of 0.5 so that a coin-flip outcome maps to zero rather than to 0.5 (Eq. \ref{eq:Cb}).

\begin{equation}\label{eq:Cb}
C_b = \frac{p_{\text{maj}} - 0.5}{1 - 0.5} = 2 p_{\text{maj}} - 1
\end{equation}

An exact tie has no majority. The implementation resolves the point estimate deterministically in such cases, but a tie gives $p_{\text{maj}} = 0.5$ and hence $C_b = 0$, so it can never cross the acceptance threshold and is routed for review rather than committed.

Low agreement on a Boolean question is a useful diagnostic. Persistent disagreement on a \texttt{NutrientFlag} often indicates a boundary case, such as whether a lightly fermented product counts as alcohol-free under the relevant definitional threshold.

\subsection{Multiple-choice Values}
\label{sec:multiple-choice-values}

Multiple-choice values generalise the Boolean case to $k$ fixed options. A typical example is a selection from a fixed set of allowed units, such as \{whole, litre, slice, gram, $\ldots$\}. The point estimate is the \emph{mode}, which is robust to scattered noise in the same way the median is robust to extreme values. The confidence score uses the mode proportion among substantive parsed responses and the same normalisation as the Boolean case, with the chance baseline adjusted from 0.5 to $1/k$ (Eq. \ref{eq:Cm}).

\begin{equation}\label{eq:Cm}
C_m = \frac{p_{\text{mode}} - 1/k}{1 - 1/k}
\end{equation}

Setting $k = 2$ recovers $C_b$ as a special case.

A supplementary diagnostic is the \emph{plurality margin} $m = p_{\text{mode}} - p_{\text{second}}$, the gap between the first- and second-ranked options. A narrow margin flags cases where the model is torn between two alternatives, for example \emph{litres} and \emph{millilitres}.

The multiple-choice estimator is a general-purpose component. In the present pipeline it is not deployed as a standalone field; the only fixed-set choice we collect is the unit of a reference quantity, and that unit is estimated jointly with its magnitude as the open-categorical pair of Section~\ref{sec:open-categorical-values}.

\subsection{Open Categorical Values}
\label{sec:open-categorical-values}

Open categorical responses can be conceptualised as the limiting case of a multiple-choice format as \(k \to \infty\), wherein the set of possible response categories becomes effectively unbounded. The \emph{reference quantity} for an ingredient is a typical example consisting of a numeric value and a unit name whose combination cannot be predefined. The point estimate is the mode, computed over whole \texttt{(magnitude, unit)} pairs rather than over each component separately. A value such as ``100 grams'' can therefore be selected only if that full pair appeared among the substantive responses. Taking the limit of the normalised expression (Eq. \ref{eq:Cm}) as $k \to \infty$ drives the chance baseline to zero, thereby isolating the unadjusted modal proportion among substantively parsed responses (Eq. \ref{eq:limit}).

\begin{equation}\label{eq:limit}
C_o = \lim_{k \to \infty} \frac{p_{\text{mode}} - 1/k}{1 - 1/k} = p_{\text{mode}}
\end{equation}

This is not a more lenient metric. In an unbounded response space, a match between two responses carries substantially more information than it does in a binary choice, because coincidental agreement is unlikely. With the common threshold $\tau = 0.90$, this requires a very high level of raw agreement.

\subsection{Optional Integer Values}
\label{sec:optional-integer-values}

The optional integer case is genuinely compound. The quantity assumes an integer value when it is defined, but it may be entirely undefined in some instances. The glycaemic index provides a canonical example: it is defined exclusively for carbohydrate-containing foods and is not applicable to substances such as olive oil or salt, which lack relevant carbohydrate content. Treating \emph{not applicable} as zero would corrupt the estimate. Discarding it entirely would lose a useful signal about applicability.

We decompose the problem into two sub-questions, each drawing on a framework already established. The first is a Boolean vote on applicability. The point estimate is \texttt{None} only when \texttt{None} responses strictly outnumber numeric ones, so an even split resolves to the numeric branch. Otherwise, the \texttt{None} responses are excluded from the analysis, and the point estimate is the median of the remaining integer-valued responses, rounded to the nearest integer. Confidence is the product of two factors. The \emph{applicability confidence} $C_{\text{app}}$ is the normalised Boolean agreement on the applicability question, against the same 0.5 baseline as Section~\ref{sec:boolean-values}. The \emph{value confidence} reuses the numerical estimator of Section~\ref{sec:numerical-values}, mapping the robust CV of the numeric responses into a bounded score and inheriting its zero-handling special cases (Eq. \ref{eq:Cval}),

\begin{equation}\label{eq:Cval}
C_{\text{val}} = \frac{1}{1 + \text{CV}_{\text{robust}}},
\end{equation}

and the composite is $C_{\text{opt}} = C_{\text{app}} \cdot C_{\text{val}}$. This multiplicative structure is low whenever either factor is low, so both questions must be resolved confidently. When the estimate is \texttt{None}, $C_{\text{val}}$ is undefined and $C_{\text{app}}$ alone applies.

The minimum sample floor is applied to the selected branch. If the estimate is numeric, at least $n_{\min}$ numeric responses are required after \texttt{None} responses have been discarded. If the estimate is \texttt{None}, at least $n_{\min}$ \texttt{None} responses are required. Domain-enforced bounds (for GI, 0-100) are imposed at the parsing stage, so out-of-range responses register as parse failures rather than inflating the spread.

\subsection{Summary}
\label{sec:estimator-summary}

Table \ref{tab:summary} summarises the framework across all five data types. For each case, there is a robust point estimator, a normalised confidence score, and a threshold-based stopping rule sitting above the shared preliminary filters. A full glossary of symbols is given in Appendix \ref{app:symbols}.

\begin{table}[ht!]
\caption{Per-field estimators and confidence scores.}\label{tab:summary}
\centering
\begin{adjustbox}{max width = 1\textwidth}
\begin{tabular}{llll}
\hline
\textbf{Data type} & \textbf{Example} & \textbf{Point estimate} & \textbf{Confidence score}\\
\hline
Numerical & nutrient ratio, cost & Median & $C_n = 1/(1 + \text{CV}_{\text{robust}})$\\
Boolean & nutrient flag & Majority vote & $C_b = 2p_{\text{maj}} - 1$\\
Multiple-choice & unit of reference quantity & Mode & $C_m = (p_{\text{mode}} - 1/k)/(1 - 1/k)$\\
Open categorical & reference quantity & Mode of complete pair & $C_o = p_{\text{mode}}$\\
Optional integer & glycaemic index & Applicability vote, then median & $C_{\text{opt}} = C_{\text{app}} \cdot C_{\text{val}}$\\
\hline
\end{tabular}
\end{adjustbox}
\end{table}

The common threshold $\tau = 0.90$ (see Table \ref{tab:stopping-controls}) is a shared default for normalised confidence scores, not evidence that all data types have the same acceptance error rate. Collection ends successfully when the confidence score crosses its threshold, or unsuccessfully when an early-exit condition is triggered. A successful result includes a value and confidence score. An unsuccessful result includes only a reason. The consistency assumption from the numerical case underlies all five strategies, while the invariant-checking layer in Section~\ref{sec:invariants} provides complementary protection against systematic errors.

\begin{table}[ht!]
\caption{Shared stopping controls used in the experiments.}
\label{tab:stopping-controls}
\centering
\footnotesize
\begin{tabular}{p{0.37\linewidth}p{0.18\linewidth}p{0.34\linewidth}}
\hline
\textbf{Shared Control} & \textbf{Experimental Value} & \textbf{Role}\\
\hline
Acceptance threshold $\tau$ & 0.90 & Required normalised confidence score\\
Minimum relevant samples $n_{\min}$ & 5 & Floor before confidence-based stopping\\
Maximum queries $n_{\max}$ & 20 & Cost cap before manual review\\
Consecutive declines & 5 & Early review trigger for repeated \texttt{DECLINE} responses\\
Parse failures & 3 & Early review trigger for schema failures\\
\hline
\end{tabular}
\end{table}

\section{Experimental Design}\label{sec:experimental-design}

\subsection{Evaluation Question}
\label{sec:evaluation-question}

The empirical evaluation investigates whether the proposed pipeline produces more usable ingredient data than a naive LLM-based collection strategy. Usability is operationalised through three measured dimensions: (i) coverage of ingredient-attribute pairs, (ii) accuracy relative to defensible reference values, and (iii) collection cost. The unit of analysis for coverage and accuracy is an ingredient-attribute pair, such as the calcium content of 100 g of cooked brown rice or the simple-sugar content of a raw apple. Cost is summarised both per ingredient and across the complete evaluation corpus.

\subsection{Compared Collection Strategies}
\label{sec:compared-strategies}

We evaluate two deployable collection strategies on overlapping data: a naive one-shot baseline and the full proposed pipeline. The saturation sampling study is not a collection strategy but an experimental instrument, described below.

The \emph{naive baseline} sends exactly one prompt per attribute, then applies the same parser and unit-normalisation layer as the full pipeline, accepting the parsed values as-is. Outcomes of \texttt{DECLINE} and \texttt{PARSE\_FAIL} are treated as unusable. This baseline is inexpensive and operationally simple, but it lacks repeated sampling, confidence thresholds, invariant checks, and any review-routing mechanism beyond parser failure.

The \emph{full pipeline} transforms an ingredient name into a complete ingredient record through the sequence in Figure~\ref{fig:pipeline}, applying statistical sampling, web-based retrieval where necessary, then tree-balancing and invariant checks, with deterministic or LLM-based repair on failure.

The \emph{saturation sampling study} (Section~\ref{sec:saturation}) underpins the comparison rather than competing in it. For each sampled ingredient-attribute item, it collects 50 repeated samples across nutrient ratios, reference quantity, glycaemic index, nutrient flags, and cost per $100g$ (\texttt{grams\_per\_reference\_quantity} was excluded as it would couple to the reference-quantity decision). Because the production pipeline normally stops early, this 50-sample regime is diagnostic rather than deployable. It is analysed retrospectively to determine when the estimators would have crossed their stopping thresholds and to quantify the variability of one-shot performance. The naive baseline's replicates are drawn directly from this matrix, each one a single sample column across all ingredient-attribute pairs.

The invariant-checking layer of Section~\ref{sec:invariants} belongs to the pipeline alone, so it is not applied to the naive baseline or the reference dataset against which both are scored. Enforcing invariants on either would blur the separation of layers that the experiment is built to isolate. This matters for the reference set in particular, whose nutrient trees do not always balance. Analogous inconsistencies have been documented in public food composition databases~\cite{ferrazdearruda2023}. Reference values are therefore left in their original, possibly unbalanced form rather than adjusted to satisfy pipeline invariants.

\subsection{Model and Prompting}
\label{sec:model-prompting}

Statistical sampling and one-shot baseline evaluations are conducted using OpenAI’s \texttt{gpt-5-mini} model via the Chat Completions API. A single, moderately capable and cost-effective model is employed because the primary objective of this work is to assess the data collection methodology rather than to perform comparative model evaluation. Investigation of robustness across multiple model architectures and configurations is deferred to future research.

Each call is a fresh, self-contained, zero-shot request to the bare model: a single user prompt with no additional context or tools. We set reasoning effort to \texttt{minimal} for sampling calls, but do not set temperature or other sampling parameters explicitly. Repeated calls therefore sample under the provider's default behaviour. Prompt templates are attribute-specific, and the following prompt (Listing \ref{list:promptTemplate}) is an example of a nutrient ratio query.

{\small
\begin{lstlisting}[frame=single,caption=Example template for a nutrient-ratio query prompt.,label=list:promptTemplate]
How much {nutrient} is in {ref_magnitude} {ref_unit} of
{ingredient}. Respond with only a number followed 
by a mass unit from {unit_list}.
\end{lstlisting}
}

\subsection{Reference Dataset}
\label{sec:reference-dataset}

The primary comparison scores both the naive baseline and the full pipeline against a manually researched \emph{reference set} (a curated collection of 30 ingredients drawn from the recipe corpus, for which ingredient-attribute values were established manually from reputable sources).

The reference set is deliberately small. Collecting it at this level of detail took several weeks of careful research, even with AI-driven deep-research tools. The design therefore extracts a strong signal from 30 carefully curated ingredients instead of a larger but weaker base. The expensive manual-research budget is spent on a defensible evaluation set, and that evidence is then used to judge whether the automated pipeline is reliable enough for larger-scale ingredient collection.

The set is fixed before analysis and spans three strata: well-documented staples, transformed or cooked ingredients, and less common ingredients where coverage is expected to be weaker. Reference values are drawn from reputable sources such as USDA FoodData Central, CoFID, peer-reviewed literature, or supplier data where appropriate. Coverage statistics are computed against the canonical ingredient-attribute grid, which is defined independently of the reference set. Pairs without a defensible reference value remain in the coverage and review-rate denominators but are excluded from accuracy comparisons.

\subsection{Statistical Analysis}
\label{sec:statistical-analysis}

We compare the naive baseline and the full pipeline against the reference set along three axes: accuracy, coverage, and cost. Both systems are scored on the same grid with the same metrics. The baseline's values come from its saturation-matrix replicates, the pipeline's from its end-to-end records.

Accuracy is assessed on nutrient flags and nutrient ratios, restricted to pairs where both the scored system and the reference set have a value. We report exact-match rate with Wilson 95\% CI for flags (a binomial proportion), and median absolute percentage error with percentile-bootstrap 95\% CI for the skewed, continuous ratio errors. 

For each eligible ingredient–attribute pair, the pipeline’s final value and a summary of the usable naive responses are evaluated against the same reference. The naive summary is the median for nutrient ratios and the majority vote for nutrient flags. Pairs are resampled with replacement using identical indices for both methods. For bootstrap resample $b \in \{1, \ldots, 10{,}000\}$, the differences are defined as:
$$
\Delta_{b,\text{nut-ratio}} =
\underset{i \in S_{b,\text{nut-ratio}}}{\operatorname{median}}
\left(\mathrm{APE}_{\mathrm{naive},i}\right)
-
\underset{i \in S_{b,\text{nut-ratio}}}{\operatorname{median}}
\left(\mathrm{APE}_{\mathrm{pipeline},i}\right),
$$

$$
\Delta_{b,\text{nut-flag}} =
\mathrm{Acc}_{\mathrm{pipeline},b}
-
\mathrm{Acc}_{\mathrm{naive},b}.
$$

Here, $i$ indexes an ingredient–nutrient pair, $S_{b,\text{nut-ratio}}$ contains the ratio-pair indices drawn with replacement for resample $b$, each APE term is the indicated method’s error for pair $i$, and $\mathrm{Acc}_{\mathrm{method},b}$ is the proportion of ingredient–flag pairs in resample $b$ for which the indicated method’s flag exactly matches the reference flag. Both $\Delta$ values are oriented so that positive values favour the pipeline. The reported $95\%$ confidence intervals comprise the $2.5$th and $97.5$th percentiles of the corresponding $\Delta$ distributions.

Coverage is measured against the canonical grid and reported as a pair of proportions that bracket the comparison: how much the pipeline produced (pipeline coverage) and how much defensible manual data exists (reference coverage). Pipeline outputs falling outside the reference-covered set are reported separately as pipeline-beyond-reference; these may be useful values missed by manual research or unsupported web-fetch results, and are not counted as accurate without adjudication.

Cost is reported as wall-clock time and OpenAI API spend per ingredient, plus a corpus total.

\section{Saturation Sampling Study}
\label{sec:saturation}

The saturation sampling study tests the statistical layer in isolation, evaluating how reliably dispersion-based estimators converge across field types and how that affects the production query budget. It is a diagnostic, not a production procedure, since the production pipeline usually stops as soon as the convergence criterion is reached. It is therefore presented separately from the head-to-head comparison in Section~\ref{sec:results}.

The study also serves a second purpose. The 50-sample response distributions it records, for a fixed set of 30 ingredients with 50 independent samples per item row, are the source from which the naive one-shot baseline's replicates are later reconstructed (Section~\ref{sec:results-accuracy}).

The ingredients were fixed before analysis and stratified as staples ($n=8$), transformed or cooked foods ($n=10$), and uncommon ingredients ($n=12$). For each ingredient, five field groups were sampled: nutrient ratios, reference quantity, glycaemic index, nutrient flags, and cost per 100g \\(\texttt{grams\_per\_reference\_quantity} was excluded as it would couple to the reference-quantity decision).

\subsection{Response Coverage}
\label{sec:saturation-response-coverage}

We define an \emph{item row} as one ingredient paired with one queried field, and a \emph{cell} as one item row at one of the 50 sample positions. The 30 ingredients across five field groups expand to 4,560 item rows: 4,200 nutrient ratios (140 nutrients $\times$ 30 ingredients), 270 nutrient flags (9 flags $\times$ 30 ingredients), and 30 each for reference quantity, glycaemic index, and cost per 100g. With 50 samples per row, the saturation matrices contain 228,000 cells in total.

\begin{figure}[ht!]
\centering
\includegraphics[width=1\linewidth]{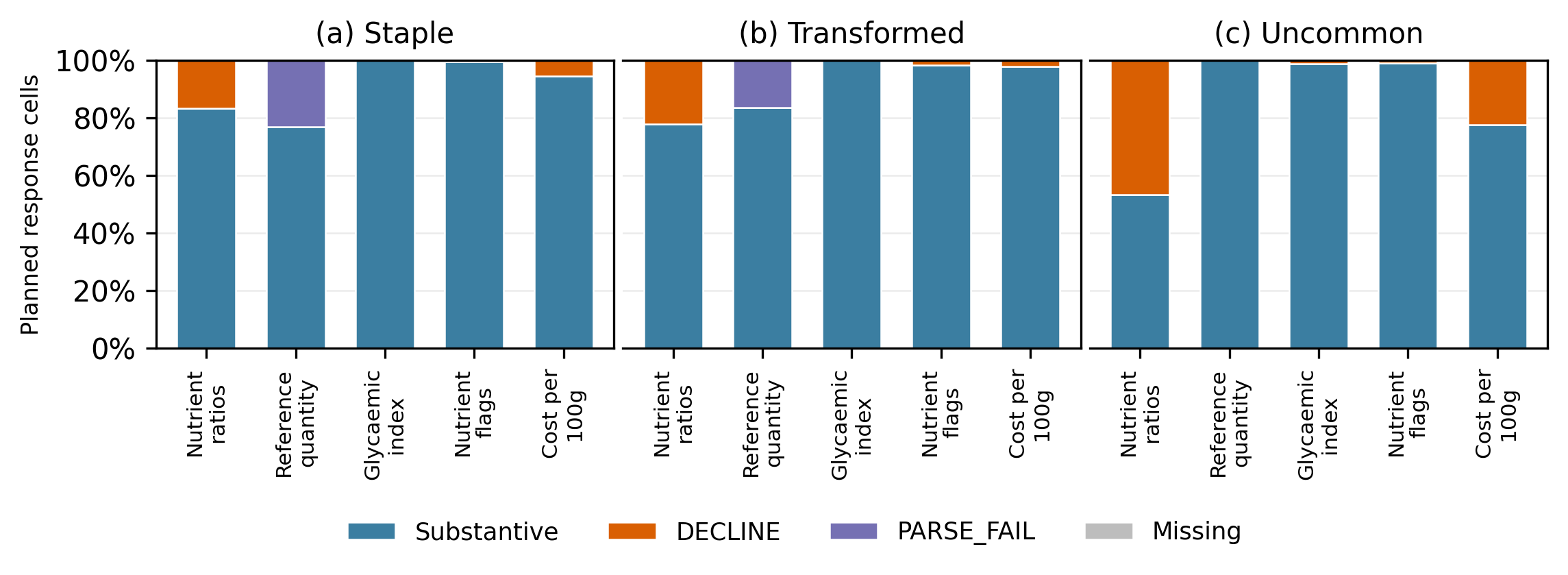}
\caption{Response coverage by field group and ingredient stratum; stacked bars show substantive, \texttt{DECLINE} and \texttt{PARSE\_FAIL} rates.}
\label{fig:coverage}
\end{figure}

All planned samples produced a recorded outcome: 163,159 substantive parsed responses (71.56\%), 64,665 explicit \texttt{DECLINE} responses (28.36\%), and 176 \texttt{PARSE\_FAIL} outcomes (0.077\%). All \texttt{PARSE\_FAIL} outcomes occurred in the reference-quantity field, where they accounted for 176 of 1,500 cells (11.73\%). The specific breakdown across the strata is shown in Figure~\ref{fig:coverage}.

\texttt{DECLINE} responses concentrated in nutrient-ratio collection (46.9\% decline on uncommon ingredients vs 16.7\% on staples) and in cost per 100g for uncommon ingredients (22.5\% decline rate). Glycaemic index and nutrient flags were near-uniformly substantive, but the unstable fields varied substantially by ingredient. Individual nutrient-ratio decline rates for uncommon ingredients ranged from approximately 33\% to over 60\%, and one ingredient saw declines in more than 90\% of cost-per-100g queries.

The reference-quantity rows did not have explicit declines, but the field-specific \texttt{PARSE\_FAIL} rate was materially higher than the global rate, mostly among staple and transformed foods. These failures are therefore reported separately from explicit \texttt{DECLINE} responses, since they indicate that no structured value could be parsed from the returned text.

\subsection{Stability and Stopping}
\label{sec:saturation-stability-stopping}

\begin{figure}[ht!]
\centering
\includegraphics[width=\linewidth]{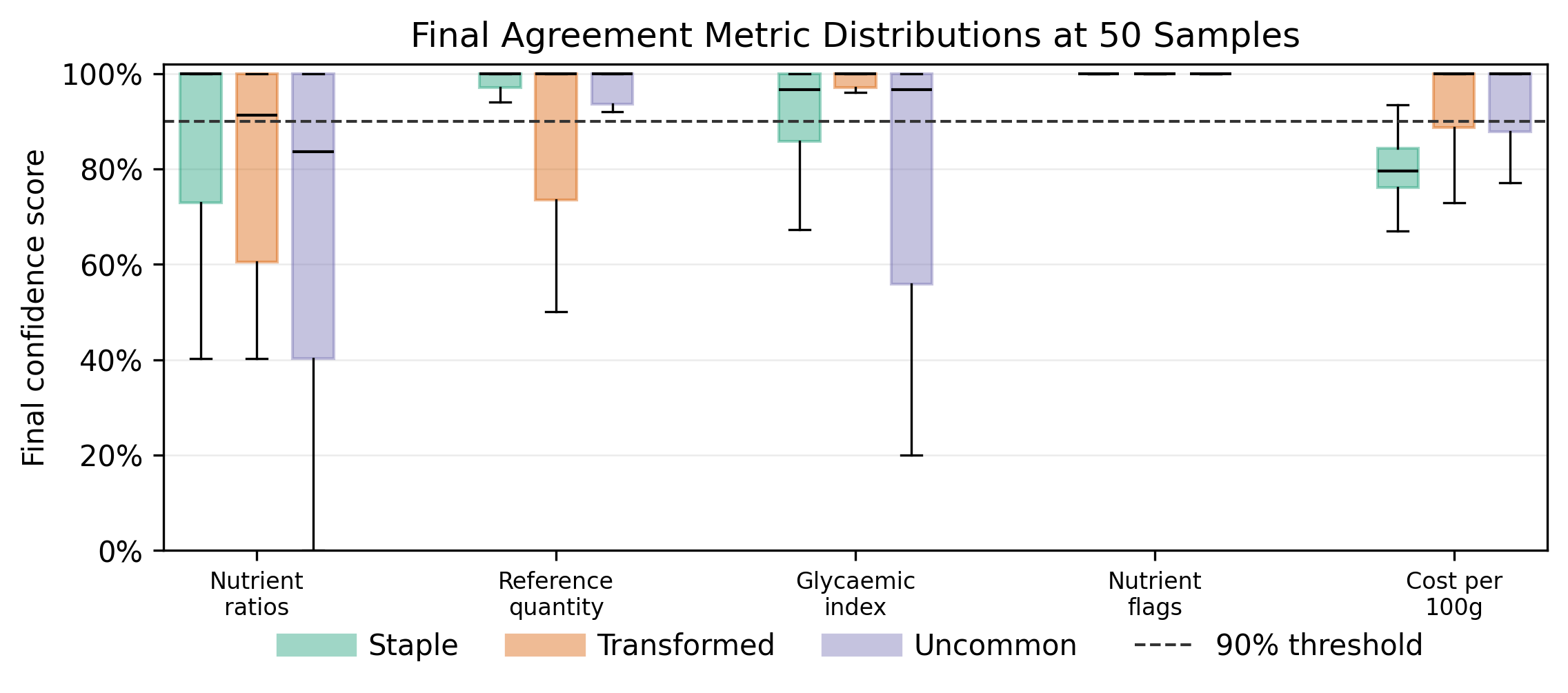}
\caption{Agreement-metric distributions at 50 samples; the dashed line marks the 0.90 acceptance threshold.}
\label{fig:final-confidence}
\end{figure}

Figure~\ref{fig:final-confidence} reports the final 50-sample confidence distribution for each field group and stratum. The 50-sample regime was chosen as a practical diagnostic budget: large enough for most fields with a stable answer distribution to settle, but still feasible to run across all item rows. Fields that remained below threshold after 50 samples are therefore treated as likely unstable under repeated sampling. The same 50-sample matrices also supply the naive baseline's replicates in Section~\ref{sec:results-accuracy}.

Nutrient flags were the most stable field group; 247 of 270 rows (91.5\%) were above the 0.90 threshold after all samples were observed. Reference quantities were also generally stable, with 24 of 30 rows (80.0\%) above threshold. Glycaemic index rows reached threshold in 22 of 30 cases (73.3\%).

The least stable fields were nutrient ratios and cost per 100g. After 50 samples, 2,200 of 4,200 nutrient-ratio rows (52.4\%) were above threshold, with lower stability for uncommon ingredients (46.3\%) than staples (63.1\%). For cost per 100g, only 15 of 30 rows (50.0\%) were above threshold. Thus, repeated sampling most readily produced stable agreement for categorical or low-ambiguity fields, while open-ended numerical fields remain more variable.

\subsection{Retrospective Stopping}
\label{sec:saturation-retrospective-stopping}

\begin{figure}[ht!]
\centering
\includegraphics[width=\linewidth]{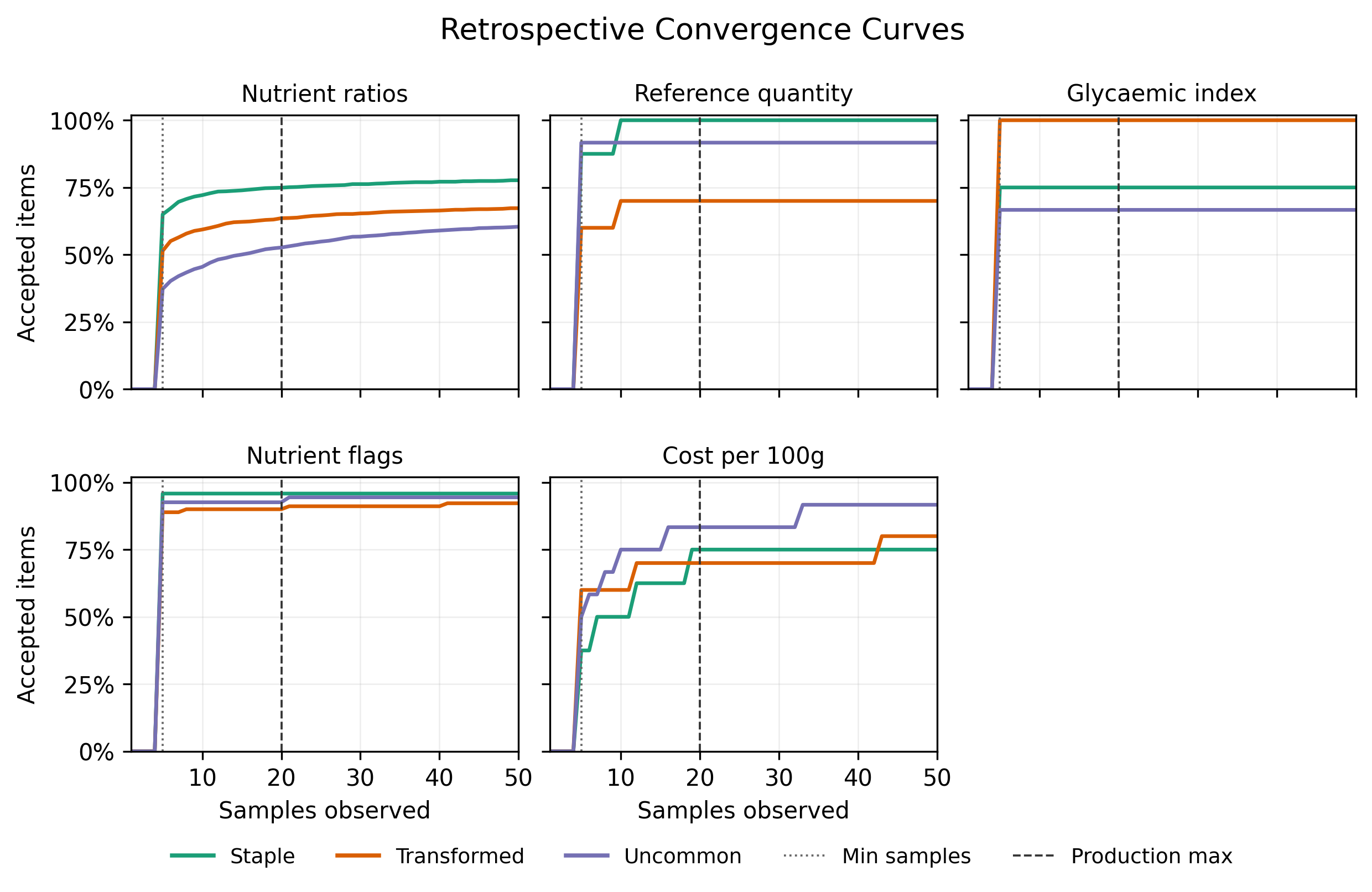}
\caption{Cumulative threshold-crossing rate as the sample prefix grows from 5 to 50.}
\label{fig:convergence-curves}
\end{figure}

Figure~\ref{fig:convergence-curves} uses the 50-sample matrices to retrospectively evaluate how the production stopping rule would have behaved within its query budget. Of the 4,560 item rows, 2,386 (52.3\%) crossed at the minimum sample count, 2,937 (64.4\%) had crossed by the production maximum of 20 queries, and 3,156 (69.2\%) crossed at some point within the 50-sample budget.

By 20 queries, nutrient flags had crossed in 250 of 270 rows (92.6\%), reference quantities in 26 of 30 (86.7\%), glycaemic index in 24 of 30 (80.0\%), and cost per 100g in 23 of 30 (76.7\%). Nutrient ratios crossed more slowly: 2,614 of 4,200 rows (62.2\%) by 20 queries and 2,827 (67.3\%) by 50. This supports a production cap of $n_{\max}=20$: most of the attainable convergence had already occurred by 20 queries, while extending nutrient ratios from 20 to 50 samples yielded only a 5.1 percentage-point gain (62.2\% to 67.3\%) at 2.5$\times$ the query cost.

Some rows clear the threshold early but later fall below it as more samples reveal dispersion. Of rows that would have been accepted within 20 queries, 619 of 2,937 (21.1\%) ended with 50-sample confidence below $0.90$, mainly in cost per 100g (43.5\% of within-20 acceptances) and nutrient ratios (22.9\%). The production threshold therefore reduces review load, but these fields need downstream invariant checks, and threshold tuning remains an important sensitivity question.

\subsection{First-convergence Timing}
\label{sec:saturation-first-convergence}

\begin{figure}[!t]
\centering
\includegraphics[width=\linewidth]{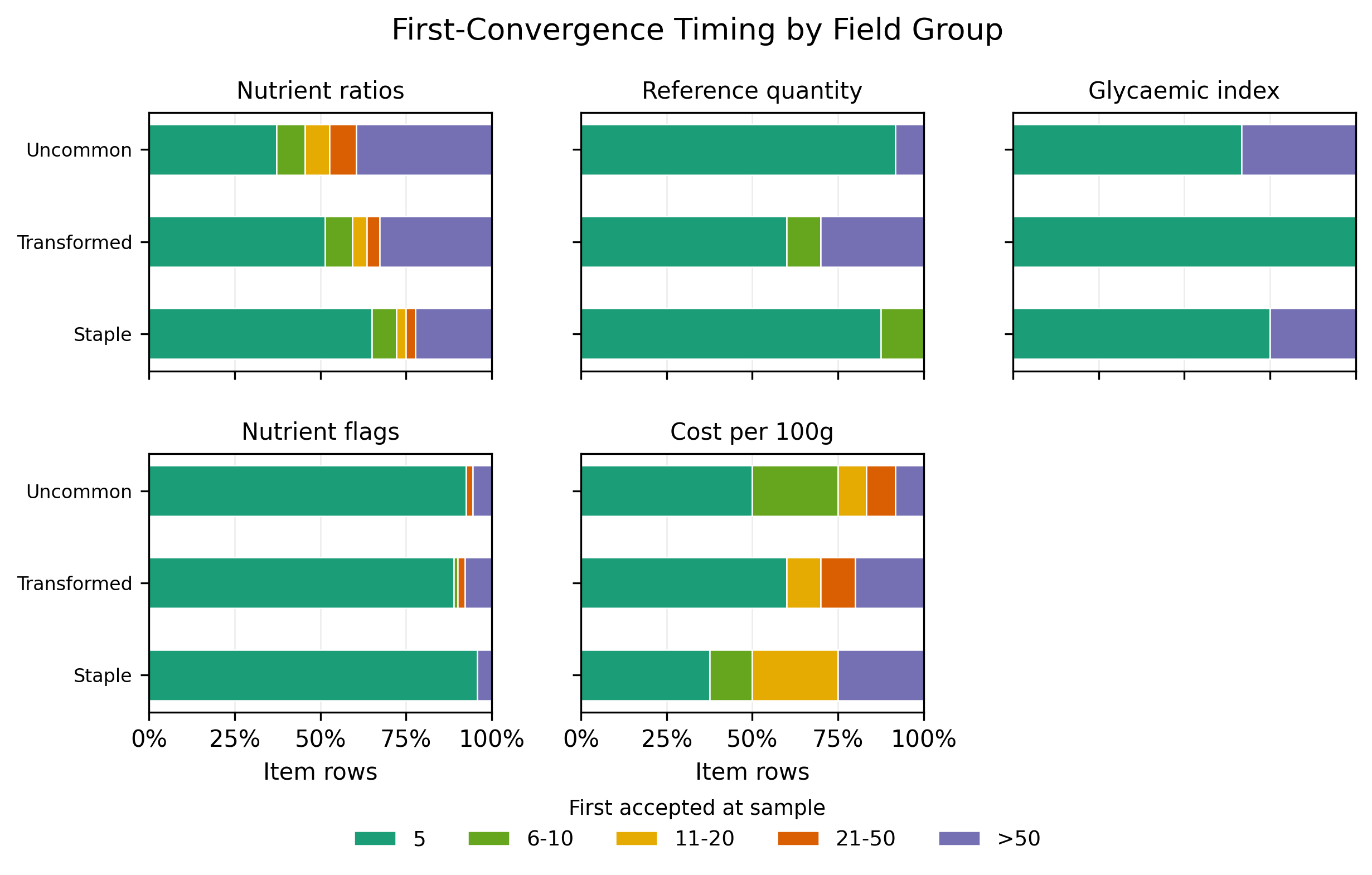}
\caption{First threshold-crossing sample within the 50-sample budget.}
\label{fig:first-convergence}
\end{figure}

Figure~\ref{fig:first-convergence} shows first-convergence timing. For all field groups, the median first accepted sample among converged rows was 5, so convergence typically occurs at the minimum sample count. The main difference is the size of the censored tail. Rows in the $>50$ bin did not reach the acceptance threshold within the 50-sample study budget.

The $>50$ category contains 1,404 rows (30.8\%). Failure to cross within 50 samples was most common for nutrient ratios, with 1,373 of 4,200 rows (32.7\%). For cost per 100g, 5 of 30 rows (16.7\%) did not cross within 50 samples; for reference quantity, 4 of 30 (13.3\%); for glycaemic index, 6 of 30 (20.0\%); and for nutrient flags, 16 of 270 (5.9\%).

Repeated querying is useful but field-dependent. As in Figure~\ref{fig:pipeline}, rows that do not converge within the production budget are not considered solved by repetition alone: they are routed to web-fetch fallback, tree balancing, invariant checks, and ultimately manual review if unresolved.

\section{Fact-Checking Ingredient Data with Invariants}
\label{sec:invariants}

The proposed pipeline first estimates each field separately through repeated LLM sampling. The invariant layer then checks whether the assembled ingredient record is internally consistent. Fields that fail to converge are routed to web-based retrieval, while invariant violations enter a repair loop: proposed repairs are applied, the record is rechecked, and the process repeats until the record satisfies the invariants or the retry budget is exhausted.

Throughout this section, a gathered \textit{zero} is treated as an absence claim rather than a small numeric estimate. The balancing layer may adjust the magnitudes of nutrients already present, but it may not change the record's presence pattern by turning zero into a positive value, or a positive value into zero. Such changes alter the ingredient's semantic identity, for example by turning an alcohol-free ingredient into one containing alcohol. Zero-crossing cases are therefore treated as factual contradictions requiring evidence-grounded repair, not as numerical imbalances for the LP to smooth away.

Table~\ref{tab:invariant-errors} lists the detectable error classes used for ingredient-level checking.

\begin{table}[ht!]
\renewcommand{\arraystretch}{1.2}
\caption{Breakdown of error types representing invariant violations.}
\label{tab:invariant-errors}
\centering
\footnotesize
\begin{tabular}{p{0.28\textwidth}p{0.64\textwidth}}
\hline
\textbf{Error type} & \textbf{Description}\\
\hline
Non-zero nutrient with zero ancestor & A child nutrient is present while an ancestor is zero, such as saturated fat > 0 but total fat = 0.\\
Child nutrients exceed parent & The sum of child \texttt{NutrientRatio} values exceeds the parent value.\\
Parent exceeds child sum & All children are defined, but their sum is smaller than the parent value.\\
Non-zero parent with all zero children & A parent nutrient is present while all explicitly defined children are zero.\\
Excluded nutrient is present & A \texttt{NutrientRatio} is non-zero despite a \texttt{NutrientFlag} saying the ingredient should not contain it.\\
False flag with true child & A parent flag is \texttt{False} while one of its child flags is \texttt{True}, such as \texttt{alcohol\_free=False} and \texttt{halal=True}.\\
Total ratios exceed 100\% & Top-level \texttt{NutrientRatio} values sum to more than 1.0.\\
\hline
\end{tabular}
\end{table}

These errors fall into two categories, each routed to a different repair path in the pipeline in Figure~\ref{fig:pipeline}.

\emph{Numerical tree imbalance errors} violate summation constraints in the nutrient hierarchy. Child nutrients may exceed their parent, the parent may exceed a fully defined child sum, a non-zero descendant may appear under a zero ancestor, or top-level ratios may exceed total mass. Small imbalances are treated as tolerable numerical discrepancies and absorbed by the deterministic tree-balancing procedure of Section~\ref{sec:tree-balancing}. Imbalances that exceed the bounded percentage tolerance, or that the linear program cannot solve under the zero-preservation constraint, escalate to LLM repair grounded in fresh web evidence, with manual review reserved for cases that web-search-grounded repair also cannot resolve.

\emph{Semantic contradictions} require a different treatment. These are not small numerical imbalances, but conflicts about what the ingredient contains. For example, a nutrient may be marked as excluded while its ratio is non-zero, a flag hierarchy may say both that a category is absent and that one of its subcategories is present, or a parent nutrient may be present while all of its defined children are zero. Moving mass around the nutrient tree cannot resolve these cases without changing the ingredient's meaning. They therefore bypass tree balancing and are routed directly to web-search-grounded LLM repair with the specific failure context, escalating to manual review only when repair cannot ground a resolution.

\subsection{Tree Balancing}
\label{sec:tree-balancing}

Nutrients are ordered in a hierarchy in which, in a complete and consistent tree, child nutrient masses sum to the parent’s mass. Exact equality is unrealistic for independently sampled LLM values, especially since the model may output \texttt{DECLINE} when evidence is weak and downstream applications must tolerate missing fields (Section~\ref{sec:introduction}). The key question is therefore how much coherent structure can be recovered from the sampled tree without making unsupported inferences.

One alternative would be to sample only nutrients with no subcomponents, such as individual sugars or specific fatty acids, and then compute every broader nutrient total by summing upward through the hierarchy. This leaf-only derivation is mathematically appealing, but empirically fragile. Leaves are often the most specialised and least documented nutrients. Missing leaves would leave large portions of the tree undefined as gaps propagate upward.

Instead, we sample values at all levels of the hierarchy and use the tree structure to reconcile only small inconsistencies ($\epsilon^* \leq 15\%$). Larger corrections, or corrections that would break the zero-preservation constraint, are treated as evidence of a factual problem rather than smoothed away.

This reconciliation can be written as a small optimisation problem: the gathered values $v_i$ define the starting point, the nutrient hierarchy defines the equality constraints, and the balanced values $x_i$ define the closest coherent tree. Because nutrients span several orders of magnitude, deviation is measured as a percentage of the original value rather than as an absolute quantity. The objective is minimax: minimise the worst-case percentage change across all nodes. Values above this bound are treated as data-quality signals that trigger escalation to web-search-grounded repair per Section~\ref{sec:invariants}.

Per the zero convention introduced at the start of the section, zero-valued nodes are fixed at zero during balancing, while non-zero nodes are allowed to move within the common percentage bound.

\begin{figure}[!t]
\centering
\includegraphics[width=\linewidth]{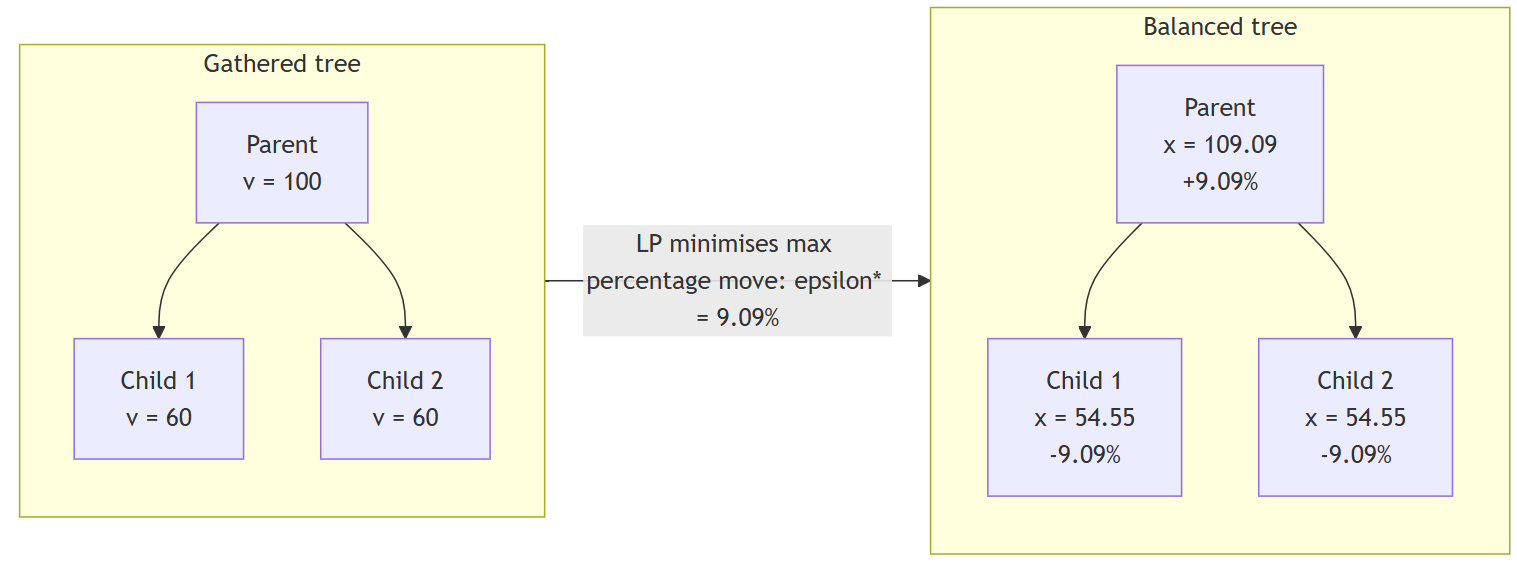}
\caption{Minimax nutrient-tree balancing example.}
\label{fig:lp-balancing}
\end{figure}

For example, suppose that an internal nutrient is gathered as 100 while its two children are gathered as 60 and 60 (Figure \ref{fig:lp-balancing}). The tree is overfilled, but the minimax solution does not hold one side fixed. Instead, it spreads the correction so that the parent rises and the children fall by the same absolute percentage change.

Let $v_i$ denote the gathered value at the nutrient node $i$ and $x_i$ its balanced counterpart, with $\epsilon$ the fractional deviation in the worst-case across all nodes and $j$ ranging over the children of $i$. The corresponding LP formulation is provided in Equation \ref{eq:LP}.

\begin{equation}\label{eq:LP}
\begin{aligned}
\min_{x,\epsilon}\quad & \epsilon \\
\text{subject to}\quad
& x_i = \sum_{j \in \mathrm{children}(i)} x_j && \forall i \in I, \\
& -\epsilon \leq \frac{x_i - v_i}{v_i} \leq \epsilon && \forall i: v_i > 0, \\
& x_i = 0 && \forall i: v_i = 0, \\
& x_i \geq 0,\; \epsilon \geq 0.
\end{aligned}
\end{equation}

Here $I$ is the set of internal nutrient nodes, such as \emph{total fat} or \emph{total carbohydrate}. The optimal value $\epsilon^*$ functions as a quantitative measure of imbalance: $\epsilon^* = 0$ indicates exact consistency, while large values trigger web-search-based repair before any manual-review fallback.

\section{Results \& Comparison with the Naive Approach}
\label{sec:results} 

This section reports the results of the experiment defined in Section~\ref{sec:experimental-design}: coverage, accuracy and cost for the proposed pipeline against the naive one-shot baseline. Pairs without a defensible reference value are included in coverage and pipeline-beyond-reference statistics, but excluded from accuracy comparisons. The pipeline yields one end-to-end record per ingredient, while the naive baseline's replicates are the 50 sample columns of the saturation matrix (Section~\ref{sec:saturation}), each parsed and unit-normalised by the same layer. The comparison grid is 4,200 nutrient-ratio pairs (140 nutrients $\times$ 30 ingredients) and 270 flag pairs (9 flags $\times$ 30 ingredients) per side, with \texttt{DECLINE} and \texttt{PARSE\_FAIL} cells treated as unusable one-shot outcomes.

\subsection{Coverage}
\label{sec:results-coverage}

Pipeline coverage for nutrient flags is 100\% (270/270): every (ingredient, flag) pair received a Boolean value, versus 90.0\% (243/270) in the reference dataset. Nutrient ratios show more variation. Table~\ref{tab:coverage} reports pipeline and reference coverage by ingredient stratum on the 4,200-pair ratio grid (140 nutrients $\times$ 30 ingredients), plus two derived metrics: \textit{reference recall} (the fraction of reference-covered pairs also covered by the pipeline) and \textit{pipeline-beyond-reference fraction} (the fraction of pipeline-covered pairs lacking a defensible reference value). 

\begin{table}[ht!]
\caption{Pipeline and reference coverage by ingredient stratum.}
\label{tab:coverage}
\centering\small
\begin{tabular}{lllll}
\hline
\textbf{Stratum} & \textbf{Pipeline} & \textbf{Reference} & \textbf{Recall} & \textbf{Beyond reference}\\
\hline
Staple ($n=8$) & 76.9\% & 63.6\% & 86.1\% & 28.8\%\\
Transformed ($n=10$) & 66.1\% & 48.3\% & 82.2\% & 39.9\%\\
Uncommon ($n=12$) & 50.4\% & 29.8\% & 71.2\% & 58.0\%\\
\textbf{Pooled ($n=30$)} & \textbf{62.7\%} & \textbf{45.0\%} & \textbf{80.8\%} & \textbf{42.1\%}\\
\hline
\end{tabular}
\end{table}

Pipeline coverage alone is uninformative because it can be inflated by emitting values without evidence. Recall and beyond-reference are the meaningful diagnostics. Recall shows that the pipeline finds nearly everything a manual researcher can defensibly source on staples and transformed ingredients, with lower recall on uncommon items matching the saturation study’s finding that the LLM declines more often on uncommon items.

The beyond-reference column is both ambiguous and central. Pooled across strata, 42.1\% of pipeline-covered ratio pairs lack any defensible reference in the manual dataset, rising to 58.0\% for uncommon ingredients. This does not show that the pipeline beats manual research, because these values are not ground-truth-checked, nor does it show fabrication, because the reference research intentionally excluded values without a strong source trail. Instead, we interpret this as an adjudication queue: the pipeline produces many values outside the manually defensible reference set, especially where reference coverage is weakest. These outputs may include useful recoveries, but they cannot be counted as accurate without further validation. Accordingly, the pipeline-reference intersection (1,525 ratio pairs and 243 flag pairs pooled) is the conservative denominator for accuracy in Section~\ref{sec:results}.

Naive replicates show consistently high flag coverage but materially lower nutrient ratio coverage. The median fraction of substantively parsed ratio responses across the 50 one-shot passes is 69.3\% (IQR [69.0\%, 69.6\%]), with the remainder consisting of \texttt{DECLINE} outcomes concentrated in uncommon ingredients. Flag responses are near-complete across replicates, with median substantive coverage of 98.9\% (IQR [98.5\%, 99.3\%]).

\subsection{Accuracy Against References}
\label{sec:results-accuracy}

Accuracy is reported separately for nutrient flags and nutrient ratios. Flags use exact-match accuracy; ratios use absolute percentage error (APE), with zero-reference cases treated as exact-zero checks. We report the pipeline’s final accepted records, the distribution across 50 naive one-shot deployments, and paired-bootstrap improvements on matched reference pairs.

\subsubsection{Nutrient Flags}
\label{sec:nutrient-flags-accuracy}

Among the 243 reference pairs covered by the pipeline, it matches 239 exactly: $98.4\%$ pooled accuracy (Wilson $95\%$ CI $[95.8\%, 99.4\%]$). Accuracy is $100\%$ for staples (65/65) and uncommon ingredients (101/101), and $94.8\%$ for transformed ingredients (73/77).

Across the 50 naive one-shot deployments, the median exact-match rate is $95.9\%$ (IQR $[95.0\%, 96.3\%]$). On matched pairs, the pipeline exceeds the naive majority-vote summaries by $2.1$ percentage points (paired-bootstrap $95\%$ CI $[-0.8, +4.9]$). Because the interval includes zero, the results show no clear difference between methods. This near-ceiling performance is expected because most flags concern common dietary facts.

\subsubsection{Nutrient Ratios}
\label{sec:nutrient-ratios-accuracy}

\begin{figure}[H]
\centering
\includegraphics[width=0.9\linewidth]{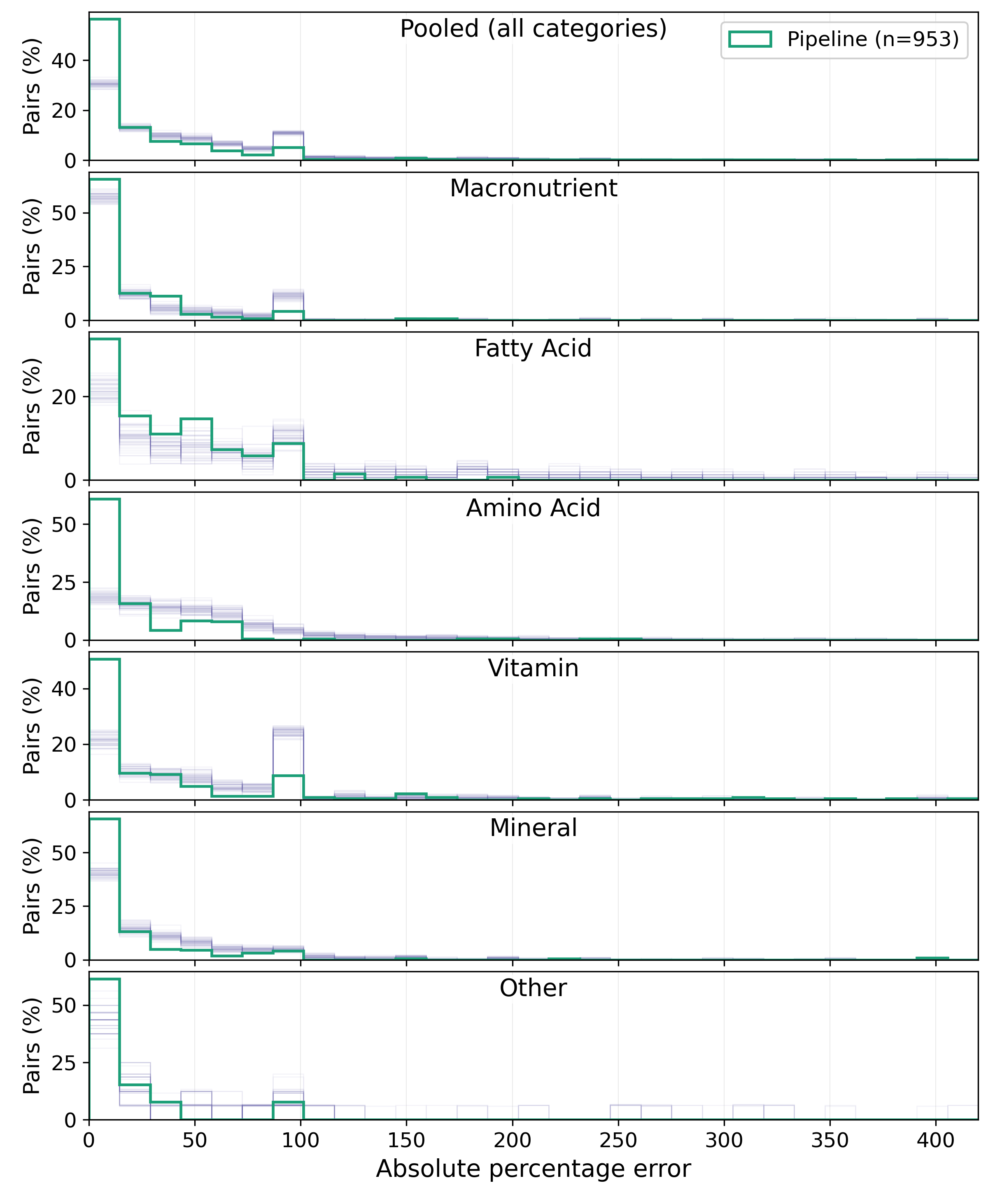}
\caption{Absolute percentage error distributions for the pipeline and 50 naive replicates. Lower panels facet by nutrient category on shared x-axis bins; median lines and IQR shaded.}
\label{fig:ape-pipeline-naive}
\end{figure}

On 953 non-zero pipeline-reference overlaps, the pipeline’s pooled median APE is $10.1\%$ (bootstrap $95\%$ CI $[8.9\%, 12.6\%]$). The 50 reconstructed naive one-shot deployments are much less accurate as standalone runs: the typical replicate has median APE $38.0\%$ (IQR $[37.6\%, 39.0\%]$; Figure~\ref{fig:ape-pipeline-naive}).

On the 948 matched non-zero pairs, the pipeline has median APE $10.1\%$ against $31.9\%$ for the median-aggregated naive summaries, an improvement of $21.8$ percentage points (paired-bootstrap $95\%$ CI $[16.3, 25.6]$).

The stratum-level pattern is not uniform. Pipeline median APE is lowest on staples at 7.0\% (bootstrap 95\% [4.7\%, 9.2\%]), highest on transformed ingredients at 19.1\% ([16.4\%, 23.5\%]), and 8.0\% on uncommon ingredients ([7.6\%, 8.8\%]). In paired-bootstrap comparisons against the naive baseline, pipeline median APE is 12.0 pp lower on staples (bootstrap 95\% CI: 6.8 to 18.9 pp lower), 26.8 pp lower on transformed ingredients (16.4 to 32.7 pp lower), and 24.7 pp lower on uncommon ingredients (14.8 to 34.8 pp lower).

The uncommon-stratum result looks better than the saturation study alone would suggest because of an overlap-selection effect. The manual researcher was able to find defensible reference values for only about 30\% of the ingredient-nutrient pairs for uncommon ingredients, and these tended to involve macronutrients and familiar minerals, where both methods performed well. The more difficult vitamins and trace minerals were mostly left unevaluated, rather than being evaluated poorly.

Of the 572 ratio pairs whose reference value is zero, the pipeline returns an exact-zero value in 96.0\% of cases (549/572). These pairs are excluded from the median APE calculation because percentage error is undefined when the reference is zero. The exact-zero agreement shows that the pipeline also performs well on semantic absences, which are not represented by the non-zero APE analysis.

Some large percentage errors are visually severe but practically modest in absolute terms, especially for trace vitamins and minerals, where doubling a very small reference quantity may still produce a negligible contribution to whole-diet totals. For example, for mature cheddar cheese, the pipeline reports vitamin B7 at 3.6 micrograms per 100g, against a reference value of $1.79$ micrograms per $100 g$. This is a 101\% absolute percentage error, but the absolute difference is only 1.81 micrograms per 100g. APE is therefore sensitive to small reference quantities. It is useful for comparing methods on the same pairs, but can overstate the practical dietary significance of the resulting errors.

\begin{figure}[ht!]
\centering
\includegraphics[width=0.8\linewidth]{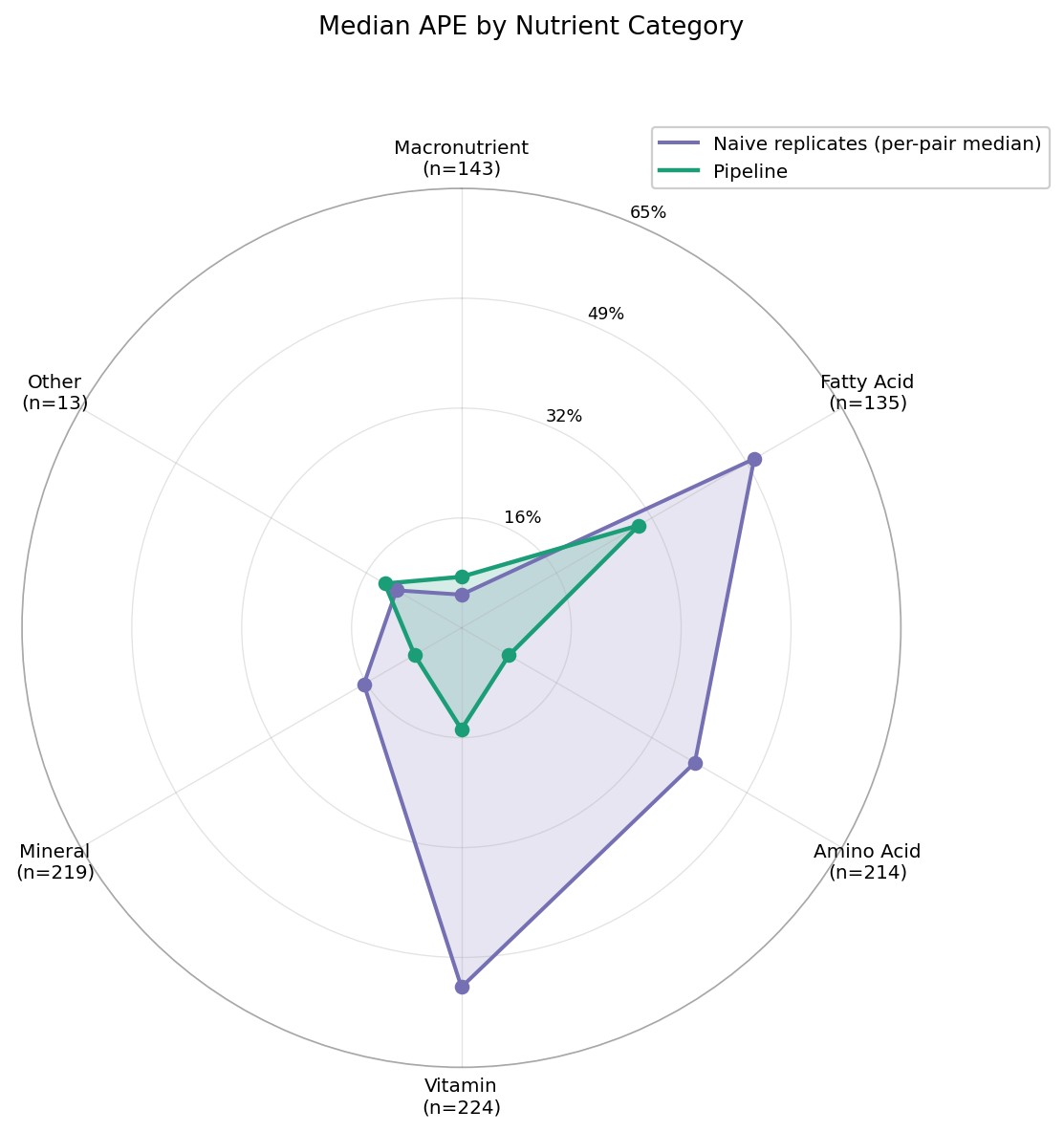}
\caption{Median APE by nutrient category on the head-to-head intersection set: pipeline (green) versus per-pair median naive baseline (purple). Axis labels show pair counts.}
\label{fig:category-apes}
\end{figure}

The accuracy gap is unevenly distributed across nutrient categories. Figure~\ref{fig:category-apes} shows median APE on the same head-to-head set of (ingredient, nutrient) pairs, split into six nutrient categories. The pipeline sits inside the naive curve on the four categories where statistical aggregation matters most. Vitamins (pipeline 15.0\% APE vs naive 53.1\%, gap 38.1 pp on $n=224$), amino acids (8.1\% vs 39.9\%, gap 31.8 pp on $n=214$), fatty acids (30.3\% vs 50.0\%, gap 19.7 pp on $n=135$) and minerals (8.0\% vs 16.7\%, gap 8.7 pp on $n=219$).

For macronutrients, the comparison reverses, and the pipeline's 7.6\% median APE is slightly higher than the naive baseline's 4.9\% in $n=143$ pairs. Macronutrients (water, protein, fat, carbohydrate, dietary fibre) are well-known enough that one-shot collection already gets them within a few percent, and the pipeline's tree-balancing layer makes small redistributions to satisfy parent-equals-sum-of-children that can move values a percentage point or two away from independent reference points.

The \textit{other} category contains 13 paired observations of compounds where the pipeline and the naive baseline are within noise of each other ($13.1\%$ vs $11.1\%$). The pipeline earns its accuracy where individual responses scatter most, where reference data itself is most sparse and variable.

\begin{figure}[ht!]
\centering
\includegraphics[width=0.9\linewidth]{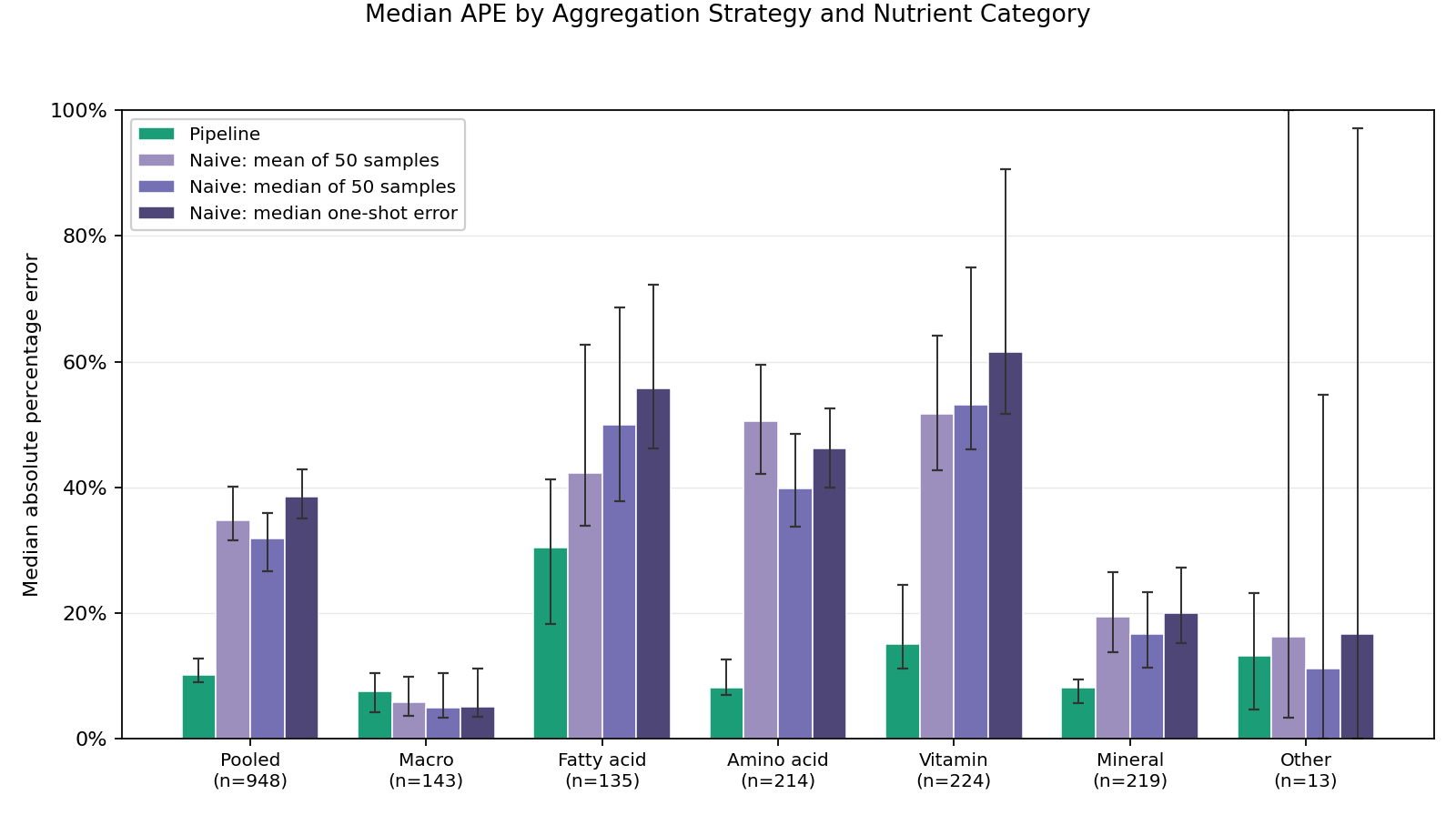}
\caption{Median APE by aggregation strategy and nutrient category on the head-to-head intersection set. Error bars show 10,000-resample percentile-bootstrap 95\% CIs.}
\label{fig:aggregation-strategies}
\end{figure}

To test whether the pipeline’s advantage could be reproduced by simple aggregation alone, Figure~\ref{fig:aggregation-strategies} compares its final estimates with three baselines constructed retrospectively from the 50 repeated responses collected for each ingredient-nutrient pair in the saturation study. Two baselines reduce the usable responses for each pair to a single estimate using either the arithmetic mean or the median, then calculate that estimate’s APE against the reference. The one-shot baseline instead calculates the APE of every usable response and takes their median, representing the error of a typical individual response for that pair. This produces one pair-level APE for each strategy, including the pipeline. The figure reports the median of these APEs across pairs, while the error bars show 95\% percentile-bootstrap confidence intervals obtained by resampling the pairs 10,000 times and recalculating the median.

On the intersection set, naive simple-mean has a pooled median APE of 34.8\%, naive-median 32.1\%, and median one-shot error 38.5\%. All three sit about 22 percentage points worse than the pipeline's 10.1\% (bootstrap 95\% CI [8.9\%, 12.8\%]) and within 7 pp of each other. This is consistent with the correlated-errors argument in Section~\ref{sec:repeat-queries}: repeated draws from the same model can share systematic error, so aggregation improves robustness but does not make the baseline competitive with the full pipeline. The mean and median baselines switch order across categories, indicating that the exact aggregation rule affects category-level errors, but not the main conclusion: simple aggregation of repeated one-shot samples remains far worse than the full pipeline.

\subsection{Cost}
\label{sec:results-cost}

The proposed pipeline required a median of 4 minutes and 50 seconds of wall-clock time and about \$1.00 in OpenAI API costs per ingredient, yielding a total runtime of roughly 2.5 hours and an aggregate cost of about \$30.00 for the 30-ingredient corpus. By contrast, a deployed naive baseline issues one query per field (about 140 fields per ingredient) at roughly \$0.001 per gpt-5-mini call, for a production cost of \$0.14 per ingredient. The baseline therefore costs about one-seventh as much per ingredient, but retains the accuracy gap reported in Section~\ref{sec:results}.

One potential path to reduce costs is to decouple evidence retrieval from judgement. Large-scale web scraping or non-LLM retrieval could collect candidate source material, which a cheaper model would then filter and normalise.

In headline terms, the pipeline spends roughly \$1 per ingredient to achieve a 22 percentage-point reduction in median APE on nutrient ratios. This appears reasonable for a one-time construction of a large ingredient corpus in a computational nutrition database, where each ingredient is reused across many recipes according to the Heap's Law projection in Section~\ref{sec:heaps}. Whereas existing platforms typically reason at the recipe level and would therefore incur this kind of cost for every new recipe, the ingredient-first approach amortises the expense across the entire recipe space.

\section{Discussion and Limitations}
\label{sec:discussion}

\subsection{Interpretation of Results}
\label{sec:discussion-interpretation}

Taken together, these findings support treating LLM-based data collection with current models as a staged validation problem rather than simply a prompting problem. Repeated sampling helps identify unstable fields and produce robust estimates, but simple aggregation of repeated responses did not reproduce the accuracy of the complete pipeline, suggesting that consensus alone is insufficient. The benefits were also strongly attribute-dependent: they were greatest for variable numerical fields, limited for near-ceiling Boolean flags, and absent for some well-known macronutrients. Finally, neither high coverage nor internal consistency establishes factual correctness, particularly for values outside the reference-covered set. The pipeline’s principal contribution is therefore its separation of estimation, confidence assessment, structural validation and review, rather than any claim that LLM outputs constitute autonomous ground truth.

\subsection{Limitations}
\label{sec:limitations}

Many wrong nutrient values do not violate any known invariant (particularly where the ingredient's attributes are incomplete). A value can be tightly agreed across repeated samples, not internally inconsistent, and still be wrong relative to external reference data. The invariant layer and reference-data comparison are necessary complements rather than substitutes for one another. Repeated sampling measures stability and invariants catch structural contradictions. Together they improve the chance of detecting errors, but they do not prove that an accepted value is correct.

The replicated one-shot baseline is also a narrow basis for comparison. It characterises variability under one model, provider endpoint and family of zero-shot prompt templates, and therefore tests whether the proposed pipeline improves on that particular deployment configuration. The evaluation does not establish whether the observed advantage generalises across models, providers, prompt designs or sampling settings.

Because the evaluation uses a curated set of 30 ingredients and resamples ingredient-attribute pairs, the reported intervals characterise performance across the reference-covered pairs in this corpus rather than population-level uncertainty for unseen ingredients.

We have not formally compared evidence-grounded repair against self-repair. The present pipeline uses newly retrieved web evidence because earlier exploratory runs suggested that recursive self-repair, based only on reported invariant violations, could lead the model to produce structurally compliant values without improving their factual grounding. That observation motivated the design, but a controlled comparison of repair regimes remains future work.

The production stopping rule accepts a field at its first threshold crossing, but the retrospective saturation analysis shows that 21.1\% of rows accepted within 20 queries ended below the threshold after 50 samples. This does not establish that the accepted values were incorrect, but it shows that measured agreement can be transient. Because later responses are unavailable once collection stops, the production pipeline cannot detect this subsequent drift. Early stopping therefore trades reduced query cost against a risk of premature acceptance, particularly for variable numerical fields. Future work should compare alternative sequential policies, such as field-specific minimum sample sizes, confirmation samples after first crossing, or repeated threshold satisfaction, evaluating them against external accuracy as well as later agreement.

Reference values should not be interpreted as exact ground truth. Some large discrepancies may arise because the pipeline and reference source represent different nutrient definitions, fractions, units or preparation states; phytosterol measurements, for example, can differ substantially according to what is included in the reported value. Even where the target is well aligned, food composition varies with cultivar, geography and preparation, while analytical methods introduce further measurement uncertainty. These factors impose a practical limit on agreement with any single reference value. Observed differences may therefore combine estimation error, target ambiguity and unavoidable reference uncertainty, and should be interpreted accordingly.

\section{Conclusion}
\label{sec:conclusion}

This study presented a quality-controlled methodology for using LLMs to populate structured ingredient-level nutritional data. Rather than assuming that generated values are inherently reliable, the pipeline treats them as uncertain estimates requiring staged validation. It combines repeated sampling and robust estimation with evidence-grounded repair, invariant checks and explicit review routing, producing auditable records whose uncertainty and internal consistency can be assessed.

On the 30-ingredient reference set, the complete pipeline substantially outperformed both one-shot collection and simple aggregation of repeated one-shot responses for nutrient ratios. Across 948 matched non-zero ingredient-nutrient pairs, it reduced median APE from 31.9\% for the median-aggregated baseline to 10.1\%, an improvement of 21.8 percentage points. The benefit was strongly attribute-dependent, however: nutrient flags were already close to ceiling performance, while no meaningful practical benefit was observed for some well-known macronutrients. Simple aggregation did not reproduce the accuracy of the complete pipeline, although the present evaluation does not isolate the contribution of each pipeline component.

Taken together, the findings support treating LLM-assisted data collection with current models as a validation workflow rather than a one-shot generation task. They do not establish that accepted outputs constitute ground truth, particularly where no defensible reference value is available. Future work should compare evidence-grounded and self-repair regimes, assess generalisability across models and prompting configurations, investigate less costly evidence-retrieval strategies, and evaluate the pipeline under the coverage and adjudication demands of a substantially larger ingredient corpus.

\appendix

\section{Symbols}

Table \ref{tab:symbols} reports and briefly describes all symbols used throughout this manuscript.

\label{app:symbols}

\begin{table}[ht!]
\renewcommand{\arraystretch}{1.2}
\caption{Symbols used in the estimator definitions.}
\label{tab:symbols}
\centering
\footnotesize
\begin{tabular}{p{0.22\textwidth}p{0.68\textwidth}}
\hline
\textbf{Symbol} & \textbf{Meaning}\\
\hline
$n$ & Total queries issued for an item\\
$d$ & Number of \texttt{DECLINE} responses\\
$f$ & Number of parse failures\\
$n_s$ & Number of substantive parsed responses\\
$n_{\text{eff}}$ & Effective parsed sample size, $n - d - f$\\
$n_{\min}$ & Minimum substantive parsed responses before confidence scoring\\
$n_{\max}$ & Maximum queries before review\\
$\tau$ & Acceptance threshold for the relevant confidence score\\
$\tilde{x}$ & Median numerical response\\
$\text{MAD}$ & Median absolute deviation from $\tilde{x}$\\
$\text{CV}_{\text{robust}}$ & Robust coefficient of variation\\
$C_n$ & Numerical confidence score\\
$C_b$ & Boolean confidence score\\
$C_m$ & Multiple-choice confidence score\\
$C_o$ & Open categorical confidence score\\
$C_{\text{app}}$ & Applicability confidence for optional integer values\\
$C_{\text{val}}$ & Numeric value confidence for optional integer values\\
$C_{\text{opt}}$ & Composite optional-integer confidence score\\
$k$ & Number of fixed categorical options\\
$p_{\text{maj}}$ & Proportion of substantive parsed responses matching the majority\\
$p_{\text{mode}}$ & Proportion of substantive parsed responses matching the modal value\\
$p_{\text{second}}$ & Proportion of substantive parsed responses matching the runner-up\\
\hline
\end{tabular}
\end{table}

\section{MAD Scaling}
\label{app:mad}

The MAD scaling constant $1.4826$ makes the median absolute deviation comparable to the standard deviation when values are normally distributed. For a normal distribution, the median absolute distance from the median is about $0.6745\sigma$, so multiplying MAD by $1/0.6745 \approx 1.4826$ estimates the usual standard-deviation scale. The method does not require nutritional responses to be normally distributed; the scaling simply puts robust spread on a familiar scale.


\begin{thebibliography}{10}
\expandafter\ifx\csname url\endcsname\relax
  \def\url#1{\texttt{#1}}\fi
\expandafter\ifx\csname urlprefix\endcsname\relax\def\urlprefix{URL }\fi
\expandafter\ifx\csname href\endcsname\relax
  \def\href#1#2{#2} \def\path#1{#1}\fi

\bibitem{li2023}
Z.~Li, et~al., Perspective: A comprehensive evaluation of data quality in nutrient databases, Advances in Nutrition (2023).
\newblock \href {https://doi.org/10.1016/j.advnut.2023.02.005} {\path{doi:10.1016/j.advnut.2023.02.005}}.

\bibitem{brinkley2025}
S.~Brinkley, J.~J. Gallo-Franco, N.~V{\'a}zquez-Manjarrez, J.~Chaura, N.~K.~A. Quartey, S.~B. Toulabi, M.~T. Odenkirk, E.~Jermendi, M.-A. Laporte, H.~E. Lutterodt, R.~A. Annan, M.~Barboza, E.~Amare, W.~Srichamnong, A.~Jaramillo-Botero, G.~Kennedy, J.~Bertoldo, J.~E. Prenni, M.~Rajasekharan, J.~de~la Parra, S.~Ahmed, The state of food composition databases: data attributes and {FAIR} data harmonization in the era of digital innovation, Frontiers in Nutrition 12 (2025) 1552367.
\newblock \href {https://doi.org/10.3389/fnut.2025.1552367} {\path{doi:10.3389/fnut.2025.1552367}}.

\bibitem{dagdelen2024}
J.~Dagdelen, et~al., Structured information extraction from scientific text with large language models, Nature Communications 15~(1) (2024) 1418.
\newblock \href {https://doi.org/10.1038/s41467-024-45563-x} {\path{doi:10.1038/s41467-024-45563-x}}.

\bibitem{zheng2023}
Z.~Zheng, O.~Zhang, C.~Borgs, J.~T. Chayes, O.~M. Yaghi, Chatgpt chemistry assistant for text mining and the prediction of mof synthesis, Journal of the American Chemical Society 145~(32) (2023) 18048--18062.
\newblock \href {https://doi.org/10.1021/jacs.3c05819} {\path{doi:10.1021/jacs.3c05819}}.

\bibitem{hua2025}
M.~Dhaliwal, A.~Hua, L.~Pullela, R.~Burke, Y.~Qin, Nutribench: A dataset for evaluating large language models in nutrition estimation from meal descriptions, in: International Conference on Learning Representations, Vol. 2025, 2025, pp. 95927--95950.

\bibitem{denisovblanch2026}
Y.~Denisov-Blanch, J.~Kazdan, J.~Chudnovsky, R.~Schaeffer, S.~Guan, S.~Adeshina, S.~Koyejo, \href{https://arxiv.org/abs/2603.06612}{Consensus is not verification: Why crowd wisdom strategies fail for llm truthfulness} (2026).
\newblock \href {http://arxiv.org/abs/2603.06612} {\path{arXiv:2603.06612}}.
\newline\urlprefix\url{https://arxiv.org/abs/2603.06612}

\bibitem{wang2023}
X.~Wang, J.~Wei, D.~Schuurmans, Q.~V. Le, E.~H. Chi, S.~Narang, A.~Chowdhery, D.~Zhou, Self-consistency improves chain of thought reasoning in language models, in: The Eleventh International Conference on Learning Representations, 2023.

\bibitem{farquhar2024}
S.~Farquhar, J.~Kossen, L.~Kuhn, Y.~Gal, Detecting hallucinations in large language models using semantic entropy, Nature 630~(8017) (2024) 625--630.
\newblock \href {https://doi.org/10.1038/s41586-024-07421-0} {\path{doi:10.1038/s41586-024-07421-0}}.

\bibitem{hall1997}
D.~L. Hall, J.~Llinas, An introduction to multisensor data fusion, Proceedings of the IEEE 85~(1) (1997) 6--23.

\bibitem{khaleghi2013}
B.~Khaleghi, A.~Khamis, F.~O. Karray, S.~N. Razavi, Multisensor data fusion: A review of the state-of-the-art, Information Fusion 14~(1) (2013) 28--44.
\newblock \href {https://doi.org/10.1016/j.inffus.2011.08.001} {\path{doi:10.1016/j.inffus.2011.08.001}}.

\bibitem{barabasi2020}
A.-L. Barab{\'a}si, G.~Menichetti, J.~Loscalzo, The unmapped chemical complexity of our diet, Nature Food 1~(1) (2020) 33--37.
\newblock \href {https://doi.org/10.1038/s43016-019-0005-1} {\path{doi:10.1038/s43016-019-0005-1}}.

\bibitem{ferrazdearruda2023}
H.~Ferraz~de Arruda, et~al., Food composition databases in the era of big data: Vegetable oils as a case study, Frontiers in Nutrition 9 (2023) 1052934.
\newblock \href {https://doi.org/10.3389/fnut.2022.1052934} {\path{doi:10.3389/fnut.2022.1052934}}.

\bibitem{vanerp2021}
M.~van Erp, et~al., Using natural language processing and artificial intelligence to explore the nutrition and sustainability of recipes and food, Frontiers in Artificial Intelligence 3 (2021) 621577.
\newblock \href {https://doi.org/10.3389/frai.2020.621577} {\path{doi:10.3389/frai.2020.621577}}.

\bibitem{bib:izzard2023}
J.~Izzard, F.~Caraffini, F.~Chiclana, Towards a software tool for general meal optimisation, Applied Intelligence 53~(7) (2023) 7751--7775.
\newblock \href {https://doi.org/10.1007/s10489-022-03935-0} {\path{doi:10.1007/s10489-022-03935-0}}.

\bibitem{bedrac2025}
L.~Kopitar, L.~Bedra{\v{c}}, L.~J. Strath, J.~Bian, G.~Stiglic, Improving personalized meal planning with large language models: identifying and decomposing compound ingredients, Nutrients 17~(9) (2025) 1492.
\newblock \href {https://doi.org/10.3390/nu17091492} {\path{doi:10.3390/nu17091492}}.

\bibitem{GJORGJEVIKJ2026101351}
A.~Gjorgjevikj, M.~Martinc, G.~Cenikj, R.~Stojanov, J.~Drole, G.~Ispirova, G.~Menichetti, N.~Ogrinc, D.~Trajanov, S.~Džeroski, B.~{Koroušić Seljak}, T.~Eftimov, Large language models in food and nutrition science: Opportunities, challenges, and the case of foodyllm, Current Research in Food Science 12 (2026) 101351.
\newblock \href {https://doi.org/https://doi.org/10.1016/j.crfs.2026.101351} {\path{doi:https://doi.org/10.1016/j.crfs.2026.101351}}.

\bibitem{Vavken_2025}
M.~P. Vavken, M.~Ogrinc, T.~Eftimov, B.~K. Seljak, \href{http://dx.doi.org/10.1109/BigData66926.2025.11401545}{Evaluation of llms in retrieving food and nutritional context for rag systems}, in: 2025 IEEE International Conference on Big Data (BigData), IEEE, 2025, p. 6833–6838.
\newblock \href {https://doi.org/10.1109/bigdata66926.2025.11401545} {\path{doi:10.1109/bigdata66926.2025.11401545}}.
\newline\urlprefix\url{http://dx.doi.org/10.1109/BigData66926.2025.11401545}

\bibitem{huang2025}
L.~Huang, W.~Yu, W.~Ma, W.~Zhong, Z.~Feng, H.~Wang, Q.~Chen, W.~Peng, X.~Feng, B.~Qin, T.~Liu, \href{https://doi.org/10.1145/3703155}{A survey on hallucination in large language models: Principles, taxonomy, challenges, and open questions}, ACM Trans. Inf. Syst. 43~(2) (Jan. 2025).
\newblock \href {https://doi.org/10.1145/3703155} {\path{doi:10.1145/3703155}}.
\newline\urlprefix\url{https://doi.org/10.1145/3703155}

\bibitem{geng2024}
J.~Geng, F.~Cai, Y.~Wang, H.~Koeppl, P.~Nakov, I.~Gurevych, A survey of confidence estimation and calibration in large language models, in: Proceedings of the 2024 Conference of the North American Chapter of the Association for Computational Linguistics: Human Language Technologies (Volume 1: Long Papers), 2024, pp. 6577--6595.

\bibitem{rousseeuw1993}
P.~J. Rousseeuw, C.~Croux, Alternatives to the median absolute deviation, Journal of the American Statistical Association 88~(424) (1993) 1273--1283.
\newblock \href {https://doi.org/10.1080/01621459.1993.10476408} {\path{doi:10.1080/01621459.1993.10476408}}.

\bibitem{heaps1978}
H.~S. Heaps, Information Retrieval: Computational and Theoretical Aspects, Academic Press, New York, 1978.

\bibitem{benz2008}
R.~W. Benz, S.~J. Swamidass, P.~Baldi, Discovery of power-laws in chemical space, Journal of Chemical Information and Modeling 48~(6) (2008) 1138--1151.
\newblock \href {https://doi.org/10.1021/ci700353m} {\path{doi:10.1021/ci700353m}}.

\bibitem{tria2018}
F.~Tria, V.~Loreto, V.~D.~P. Servedio, Zipf’s, heaps’ and taylor’s laws are determined by the expansion into the adjacent possible, Entropy 20~(10) (2018).
\newblock \href {https://doi.org/10.3390/e20100752} {\path{doi:10.3390/e20100752}}.

\end{thebibliography}
\end{document}